\documentclass[12pt]{article}

\usepackage{bbold}
\usepackage{amssymb}
\usepackage{slashed}
\usepackage{graphics}
\usepackage{youngtab}
\usepackage{amsmath}
\usepackage{shadow}

\declareslashed{}{/}{.1}{0}{E}

\def\ss{\scriptstyle}

\def\N{{\cal N}}

\def\wt#1{\widetilde{#1}}

\def\d{{\bf d}}

\def\be{\begin{equation}}
\def\ee{\end{equation}}
\def\beq{\begin{equation}}
\def\eeq{\end{equation}}
\def\bea{\begin{eqnarray}}
\def\eea{\end{eqnarray}} 
\def\beqa{\begin{equation}\begin{array}{l}}
\def\eeqa{\end{array}\end{equation}}
\def\eqn#1{(\ref{#1})}
\def\eqref#1{eq.~(\ref{eq:#1})}

 \def\G{{\it\Gamma}}

\def\nn{\nonumber}

\def\N{{\bf N}}
\def\TR{{\bf tr}}
\def\G{{\bf g}}
\def\DIV{{\bf div}}
\def\GRAD{{\bf grad}}

\def\Nt{{\cal N}}
\def\Tt{{\cal T}}
\def\c{{\hspace{.3mm}\bf c\hspace{.3mm}}}
\def\Real{\mathbb R}

\newcommand{\bm}[1]{\mbox{\boldmath $ #1 $}}

\begin{document}

\thispagestyle{empty}

\vspace{-.2cm}
\setcounter{footnote}{0}
\begin{center}
{\Large
 {\bf Supersymmetric Quantum Mechanics and \\[3mm]
                       Super-Lichnerowicz Algebras 
 }\\[4mm]

 {\sc \small K.~Hallowell and A.~Waldron\\[2mm]}

 {\em\small Department of Mathematics, University of California,
            Davis CA 95616, USA\\ 
            {\tt hallowell,wally@math.ucdavis.edu}}\\[2mm]
}

\bigskip

{\it Dedicated to the memory of Tom Branson}\\[2mm]

\bigskip

{\sc Abstract}\\
\end{center}

{\small
\begin{quote}
\bigskip

We present supersymmetric, curved space, quantum mechanical models
based on deformations of a parabolic subalgebra of $osp(2p+2|Q)$.
The dynamics are governed by  a spinning particle action
whose internal      coordinates are
Lorentz vectors labeled by the fundamental representation
of $osp(2p|Q)$.
The states of the theory are tensors or spinor-tensors on the curved background 
while conserved charges correspond to the various differential geometry
operators acting on these. The Hamiltonian generalizes Lichnerowicz's
wave/Laplace operator. It is central, and the models are supersymmetric 
whenever the background is a symmetric space, although there is an $osp(2p|Q)$
superalgebra for any curved background.
The lowest purely bosonic example $(2p,Q)=(2,0)$ corresponds to a 
deformed Jacobi group and describes Lichnerowicz's original algebra of
 constant curvature, differential geometric operators acting on 
 symmetric tensors. The case $(2p,Q)=(0,1)$ is simply the ${\cal N}=1$
 superparticle whose supercharge amounts to the Dirac operator
 acting on spinors. The $(2p,Q)=(0,2)$ model is the ${\cal N}=2$ supersymmetric
 quantum mechanics corresponding to differential forms. (This latter pair of models
 are supersymmetric on any Riemannian background.) 
 When $Q$ is odd, the models apply to spinor-tensors. The $(2p,Q)=(2,1)$
 model is distinguished by admitting a central Lichnerowicz-Dirac operator when the background
 is constant curvature. 
 The new supersymmetric models  are novel in that the Hamiltonian is not
 just a square of super charges, but rather a sum of commutators of supercharges and
 commutators of bosonic charges. 
 These models and superalgebras are a very useful tool for
 any study involving high rank tensors and spinors on manifolds.

\bigskip

\end{quote}
}

\newpage
\setcounter{page}{1}
%%%%%%%%%%%%%%%%%%%%%%%%%%%%%%%%%%%%%%%%%%%%%%%%%%%%%%%%%%%%%%%%%

\vfill

\tableofcontents

\vfill

\newpage

\section{Introduction}

Since the early 1980's it has been clear that supersymmetric quantum mechanical 
models are deeply related to geometry. In particular, Alvarez-Gaum\'e and Witten
computed gravitational anomalies and Pontryagin classes by identifying the Dirac operator
on a Riemannian manifold with the supercharge of an ${\cal N}=1$ supersymmetric quantum mechanics~\cite{Alvarez-Gaume:1983ig}. In an application to Morse theory, Witten realized that the
supercharges of  ${\cal N}=2$ supersymmetric quantum mechanics corresponded to 
the exterior derivative~$\d$  and codifferential $\bm{\delta}$~\cite{Witten:1982im}. In these models
the Hamiltonian corresponds to the Laplace operator, and states are spinors or differential forms,
respectively. Increasing the number of supersymmetries to ${\cal N}=4$ requires that the background manifold be K\"ahler~\cite{Zumino:1979et} and the supersymmetry charges are now the Dolbeault
operators. In particular the $sl(2,{\mathbb R})$  Hodge-Lefschetz algebra of the de Rham 
cohomology of a K\"ahler manifold amounts to the algebra of conserved charges of
the ${\cal N}=4$ supersymmetric quantum mechanics~\cite{Witten:1982df,Figueroa-O'Farrill:1997ks}.

If instead of studying differential forms on a manifold $M$, but rather symmetric tensors, there is
also an extremely useful algebra of the gradient and divergence operations. In particular, Lichnerowicz observed that for constant curvature backgrounds one could introduce a wave operator that generalized
the Laplacian and (in some sense) commuted with the action of the divergence and gradient
operators on symmetric tensors~\cite{Lichnerowicz:1961}. 
(His original motivation was a computation of the spin~2 
massive propagator on these spaces.) In a recent study of higher spins in constant curvature
manifolds we found the algebra of the following operators on symmetric tensors\footnote{The flat version of this algebra has been studied in the context of conformally flat manifolds in~\cite{Lecomte}.}~\cite{Hallowell}:

\vspace{.4cm}
\begin{tabular}{c|c|c}
\hline
{\it Index}&{ ${\bf N}$} &
Counts indices. $\phantom{S^{S^{S^S}}}\!\!$\\[2mm]
{\it Trace}&{ ${\bf tr}$} &
Traces
a pair of indices.\\[2mm]
{\it Metric}&{ ${\bf g}$}  &
Multiplies by the metric and symmetrizes.\\[2mm]
{\it Divergence }&{${\bf div}$} &
The covariant divergence.\\[2mm]
{\it Gradient}&{ ${\bf grad}$} &
Symmetrized covariant derivative. \\[2mm]\hline
\end{tabular}
\vspace{4mm}

\noindent
The result was a deformation of the Jacobi group (a central extension of
$sl(2,{\mathbb R})\ltimes {\mathbb R}^2$)
and  in this paper we extend it to an arbitrary symmetric space. In particular, unlike
differential forms for which the {\it anticommutator} of the exterior derivative and codifferential 
(morally -- gradient and divergence) yield the {\it form} Laplace operator
\be
\{ \d,\bm{\delta}\} \ = \ \square\,  ,
\ee
for symmetric tensors, it is the {\it commutator} that does so
\be
[{\bf div},{\bf grad}]=\Delta-{\bf g}\, {\bf tr} +{\bf N}({\bf N}+{\rm dim}_M-2)\, .\label{divgrad}
\ee
Here ${\bf g}\, {\bf tr} -{\bf N}({\bf N}+{\rm dim}_M-2)$ is the Casimir of the $sl(2,{\mathbb R})$
algebra generated by $(\G, \N,\TR)$ and the central Lichnerowicz wave operator equals
\be
\Box=\Delta+{\bf g}\, {\bf tr} -{\bf N}({\bf N}+{\rm dim}_M-2)\, .
\ee
The natural question, posed and answered here, is whether there exists a Lichnerowicz/Lefschetz
type algebra applicable to both differential forms and symmetric tensors,  or indeed, for the most
general tensors and spinor-tensors on a Riemannian manifold. We answer this question
in the affirmative and find that there exists a central Lichnerowicz wave operator $\square$
whenever the manifold $M$ is a locally symmetric space, namely its curvature is
covariantly constant
\be
D_\kappa R_{\mu\nu\rho\sigma}=0\, .
\ee
The operator $\square$ also exists for general Riemannian manifolds and always 
commutes with an $osp(2p|Q)$ subalgebra of operators generalizing $\{\G,\N,\TR\}$.

We construct these ``super-Lichnerowicz'' algebras from the conserved charges of a supersymmetric quantum mechanical 
model whose states are tensors and tensor-spinors on~$M$. The model itself is a simple generalization of the ${\cal N}=2$ supersymmetric quantum mechanics. The action is just
\be
S=\frac12\int dt \left\{\dot x^\mu g_{\mu\nu} \dot x^\nu-iX^\mu\cdot \frac{DX_{\mu}}{dt}+\frac14 R_{\mu\nu\rho\sigma}\ 
X^\mu\cdot X^\nu\ 
X^\rho \cdot X^\sigma
\right\}\, ,
\ee
or in other words, a generalized spinning particle model~\cite{Brink:1976sz,Rietdijk:1989qa,Gibbons:1993ap}.
The fields $x^\mu(t)$ are the worldline imbedding coordinates, while $X^\mu(t)$
are a collection of both bose and fermi fields transforming under the fundamental 
representation of the superalgebra $osp(2p|Q)$ for which the dot ``$\cdot$'' denotes the 
invariant inner product. Supersymmetry requires the background to be a locally symmetric
space. Gauged models of this type where $p=0$ and the fields $X^\mu$ are purely fermionic
are known to describe massless higher spin particles\footnote{Recently, the one-loop
quantization of these models has also been studied~\cite{Bastianelli}.}~\cite{Gershun}.
(When $(2p,Q)=(2,0)$, the fields $X^\mu$ carry an $sp(2)$ index
corresponding to the  spinor and conformal
vector oscillators first introduced by
Labastida and Vasiliev to describe higher spin 
fields~\cite{Labastida:1987kw,Vasiliev:1988xc,Vasiliev:1990en}.) 

The conserved charges of the model correspond to both commuting and anti-commuting
generalizations of the gradient and divergence (alias exterior derivative and codifferential)
operators, along with a Lichnerowicz wave operator (the Hamilitonian), and operators
that trace over indices, count indices of a given type (form or totally symmetric), multiply by metric tensors.
and change indices from one type to another. Taking the parameters $(2p,Q)$ completely general
lets us describe arbitrary tensors and spinor-tensors. The interpretations for some specific values 
are listed below
\begin{itemize}
\item[{\tiny\framebox{(0,1)}}]
${\cal N}=1$ supersymmetric quantum mechanics -- describes the 
Dirac operator acting on spinors.
\item[{\tiny\framebox{(0,2)}}]
${\cal N}=2$ supersymmetric quantum mechanics -- describes the 
exterior derivative acting on differential forms.
\item[\!{\tiny\framebox{(2,0)}}] This model is purely bosonic and yields 
Lichnerowicz's original construction -- states are symmetric tensors. The conserved charge algebra
is the ``symmetric algebra'' of~\cite{Hallowell}.
\item[\!\!{\tiny\framebox{(2,1)}}] A model of totally symmetric spinor-tensors
({\it i.e.}, spinors with an arbitrary number of totally symmetric vector indices), which means
there are charges corresponding to gamma-traces (contracting vector and spinor indices 
using Dirac matrices).
This algebra
was first  encountered in a study of partially massless fermionic fields in~\cite{Deser:2001pe}
and sytematized in~\cite{Hallowell}. (It has also appeared in 
the computation of massive and partially massless fermionic actions~\cite{Metsaev:2006zy}.)
The theory is distinguished by possessing a ``Lichnerowicz--Dirac'' operator -- 
a modification of the Dirac operator that {\it commutes} with all other charges
when  the background is constant curvature (but {\it not} in general symmetric spaces).

\item[\!\!{\tiny\framebox{(2p,0)}}\!] A purely bosonic model, but  states are now ``multi-symmetric tensors''.
There are $p$ different gradient and divergence operators along with operators that remove or add
pairs of indices or change index types.

\item[\!\!{\tiny\framebox{(0,2q)}}\!]
States are multi-forms (indices are grouped into antisymmetric subsets, {\it e.g.}, 
the Riemann tensor
is a ``bi-form''). There are now $q$ distinct exterior derivatives and codifferentials. Algebras
of this type have been studied by Hull and Medeiros in the context of ``exotic'' higher spin gauge
theories~\cite{deMedeiros:2003dc} and a mathematical analysis was given in~\cite{Senovilla,Bekaert:2006ix}.

 \end{itemize}

\noindent
For $(2p,Q)=(0,1)$ or $(0,2)$, namely the ${\cal N}=1,2$ supersymmetric models, 
the algebra of conserved charges is a finite dimensional superalgebra for any background
Riemannian manifold. The same
is true whenever the target  manifold $M$ is flat but $(2p,Q)$ are arbitrary. In general backgrounds, 
we obtain a deformation of 
the finite dimensional Lie superalgebra appearing in the flat case. 
The simplest example is the $(2,0)$ model in a constant curvature background 
subject to the algebra~\eqn{divgrad}.
The Lichnerowicz wave operator~$\square$ is
central but the $sl(2,{\mathbb R})$ Casimir  does not commute with the 
doublet $({\bf grad},{\bf div})$. Instead, further commutators yield an infinite series of
higher operators (involving increasing powers of the ``oscillators'' $X^\mu$). 
In this case there exists a reformulation of the resulting infinite dimensional 
Lie algebra as a rather simple associative algebra, obtained by including a certain 
square root of the Casimir operator~\cite{Hallowell}. This is not a direction 
we pursue in this paper, but there is a simple characterization of the Lie 
superalgebra being deformed as a parabolic subalgebra ${\mathfrak p}$ of $osp(2p+2|Q)$.
This parabolic superalgebra can be viewed as a (supersymmetric) generalization of the Jacobi
group $G^{J}=(Sl(2,{\mathbb R})\ltimes {\mathbb R}^2)\times R^*$.

Our results are organized as follows: In section~\ref{model}, we briefly review $osp(2p|Q)$ superalgebras and write down our supersymmetric quantum mechanical model. Then we quantize it
and determine its Hilbert space, conserved charges, and their algebra in section~\ref{quantum}.
Section~\ref{geometry} is concerned with the application of the model to geometry.
There we discuss each of the special examples itemized above, followed by the most general case
and also the deformed parabolic
superalgebra of differential geometric operators on $M$.
(A hardy reader can skip the examples and study sections~\ref{superlich} and~\ref{totalalg} directly.) 
Our conclusions list various physical
and mathematical applications of our theory as well as speculations on its generalizations,
especially to manifolds with additional structures, such as a K\"ahler one, a possible associative
algebraic reformulation of our algebra, the {\it r\^ole} of the Jacobi group and novel Casimir operators, applications to higher spins and gauged versions of the model, extensions of
the algebra to $osp(2p+2|Q)$ by including the inverse Lichnerowicz wave operator, and finally
the model's quantum mechanical spectrum.
\begin{center}
\shabox{\begin{tabular}{c}{\it Tom Branson knew of Lichnerowicz's work.} \\ {\it He
may have liked this paper, so it's dedicated to him.} \end{tabular}}
\end{center}

\section{The Model}

\label{model}

\subsection{Orthosymplectic Superalgebras}

The Lie superalgebra $osp(2p| Q )$ is defined by even supermatrices 
\be
\lambda=\left(\begin{array}{c|c}A&B\\\hline C&D\end{array}\right)\, ,
\ee
subject to the orthosymplectic condition
\be
\lambda J = -J \lambda^{st}\, ,\label{osc}
\ee
where the supertranspose is defined by
\be
\left(\begin{array}{c|c}A&B\\\hline C&D\end{array}\right)^{\!\!st}\equiv
\left(\begin{array}{c|c}\!A^t\!\!&C^t\\\hline \!-B^t &D^t \end{array}\right)\, .
\ee
The invariant tensor/metric $J$ is given by
\be
J=\left\{
\begin{array}{cc}
\left(\begin{array}{cc|cc}&-{\bf1}_{p\times p}&&\\[1mm]
                                               {\bf 1}_{p\times p}&&&\\\hline
                                               &&& {\bf 1}_{q\times q}\\[1mm]
                                               &&{\bf 1}_{q\times q}&\end{array}\right) &  Q =2q \mbox{ even}\, ,\\[20mm]
\left(\begin{array}{cc|ccc}&-{\bf1}_{p\times p}&&&\\[1mm]
                                               {\bf 1}_{p\times p}&&&&\\\hline
                                               &&& {\bf 1}_{q\times q}&\\[1mm]
                                               &&{\bf 1}_{q\times q}&&\\[1mm]
                                               &&&&1\end{array}\right) &  Q =2q+1 \mbox{ odd}\, .
                                               \end{array}
\right.
\ee
The condition~\eqn{osc} is more simply formulated in terms of $\Lambda\equiv \lambda J$,
\be
\Lambda^{\! st} = - J^2 \Lambda\, ,\qquad J^2={\rm diag}\Big(\overbrace{-1,\ldots,-1}^{2p \ {\rm times}},
\overbrace{1,\ldots,1}^{ Q  \ {\rm times}}\Big)\, ,
\ee
whose solution is
\be
\Lambda\equiv (\Lambda^{\alpha\beta})
=\left(
\begin{array}{c|c}
\lambda_S&\epsilon\\\hline
\epsilon^t&\lambda_A
\end{array}
\right)\, ,\label{parameters}
\ee 
with $\lambda_S$ and $\lambda_A$ being  $2\times2$ symmetric and antisymmetric
bosonic matrices while $\epsilon$ is a $2\times 2$ fermionic matrix.
The superindices $\alpha,\beta,\ldots$ run over $2p$ bosonic and $ Q $ fermionic 
values. Note that transformations
\be
\delta X_\alpha = J_{\beta\alpha} \Lambda^{\gamma\beta} X_\gamma\, ,\label{osptr}
\ee
leave invariant the orthosymplectic inner product
\be
X\cdot Y\equiv - X_\alpha J^{\beta\alpha} Y_\beta\, .
\ee
In addition to time translation invariance, 
our model will enjoy $ Q (2p+1)$ supersymmetries as well as bosonic
$sp(2p)\ltimes{\mathbb R}^{2p}$ and $so([Q/2],[Q/2+1/2])$ symmetries. These will correspond to
$osp(2p|Q)$ and its fundamental representation ${\mathbb R}^{2p|Q}$.
% ``internal'' supersymmetries 
%corresponding to the Grassmann      parameters $\epsilon$ in the matrix $\Lambda$ 
%in~\eqn{parameters} and generators $f_{\alpha\beta}$
%plus ``traditional'' supersymmetry generators $v_\alpha$ belonging to the
%$osp(2p| Q )$ fundamental representation~${\mathbb R}^{2p| Q }$.
The $osp(2p|Q)$ generators obey the Lie superalgebra 
 \be
 [f_{\alpha\beta},f_{\gamma\delta}\} =4J_{(\beta(\gamma}f_{\alpha]\delta]}\, ,
\ee 
where the superbracket $[\cdot,\cdot\}$ is a commutator unless both entries
are femionic, in which case it equals the anticommutator. We will also often
need the quadratic $osp(2p|Q)$ Casimir
\be
{\bf c}=\frac12 J^{\beta\alpha}f_{\alpha\gamma}J^{\delta\gamma}f_{\delta\beta}\, .
\label{casimir}
\ee
It commutes with all the $osp(2p|Q)$ generators $f_{\alpha\beta}$.

\subsection{Orthosymplectic Spinning Particle}

The field content of our quantum mechanical theory 
consists of  the embedding coordinates $x^\mu$,
\be
x^\mu: {\mathbb R}\longrightarrow M\, ,
\ee
taking values in a $({\rm dim}_M\!=\!d)$-dimensional Riemannian target manifold
$(M,g_{\mu\nu})$,~along with $2p+Q$ additional fields $X^\mu$ 
describing spinning degrees of freedom 
 \be X^\mu_\alpha=
 \left\{
 \begin{array}{cc}
 \Big((\varphi_{1,\ldots, p}^\mu)^*,\varphi_{1,\ldots,p}^\mu,(\psi_{1,\ldots, q}^\mu)^*,\psi_{1,\ldots, q}^\mu\Big)\, ,
 &  Q =2q \mbox{ even}\, ,\
 \\[4mm]
 \Big((\varphi_{1,\ldots, p}^\mu)^*,\varphi_{1,\ldots,p}^\mu,(\psi_{1,\ldots, q}^\mu)^*,\psi_{1,\ldots, q}^\mu,\psi^\mu\Big)\, , & Q =2q + 1\mbox{ odd}\, .
 \end{array}
 \right.\ee 
The $\varphi$'s are bosonic and $\psi$'s fermionic and together $X^\mu_\alpha$
transforms as the fundamental representation of $osp(2p, Q )$ (as in~\eqn{osptr}). 
The action is the natural generalization
of the ${\cal N}=2$ real supermultiplet
\be
S=\frac12\int dt \left\{\dot x^\mu g_{\mu\nu} \dot x^\nu+iX^\mu_\alpha\,J^{\beta\alpha} \frac{DX_{\beta\mu}}{dt}+\frac14 R_{\mu\nu\rho\sigma}\ 
X^\mu_\alpha J^{\beta\alpha} X_\beta^\nu\ 
X^\rho_\gamma J^{\delta\gamma} X_\delta^\sigma
\right\}\, .\label{action}
\ee
As already mentioned, there are clearly 
%$\{2(p+ Q /2)^2+p- Q /2\}$ $osp(2p| Q )$ 
$p(2p+1)+\frac12 Q(Q-1)$ bosonic and $2pQ$ fermionic
internal 
symmetries 
obtained by $X^\mu_\alpha$ transformations
\be
\delta X^\mu_\alpha =J_{\beta\alpha} \Lambda^{\gamma\beta} X^\mu_\gamma\, .\label{internal}
\ee
%The supermatrix $\Lambda_{\alpha\beta}$ 
%obeys the supersymplectic condition
%\be
%J^{\alpha\gamma}\Lambda_{\delta\gamma}J^{\beta\delta}
%+J^{\alpha\gamma}\Lambda_{\gamma\delta}J^{\delta\beta}=0\, ,
%\ee
%so that 
%has the form
%\be
%\Lambda
%=\left(
%\begin{array}{c|c}
%\lambda_S&\epsilon\\\hline
%\epsilon^t&\lambda_A
%\end{array}
%\right)\, ,
%\ee 
%where $\lambda_S$ and $\lambda_A$ are $2\times2$ symmetric and antisymmetric
%bosonic matrices while $\epsilon$ is a $2\times 2$ fermionic matrix.
These hold for any background metric $g_{\mu\nu}$ and we discuss their interpretation later.
Less trivial are $Q$ further  supersymmetries plus their $2p$ bosonic ``partner'' symmetries given by
\bea
\delta x^\mu &=& i X_\alpha^\mu \varepsilon^\alpha \, ,
\nn\\[2mm]
{\cal D} X^\mu_\alpha &=&\ \dot x^\mu \varepsilon_\alpha\, . \label{susies}
\eea 
In the above, ${\cal D}$ and $D/dt$ are the covariant variation and worldline derivatives
defined on vectors $v^\mu$ by
\bea
{\cal D}v^\mu&\equiv&\delta v^\mu + \Gamma^\mu_{\rho\sigma}\delta x^\rho v^\sigma \ = \
\delta x^\rho D_\rho v^\mu\, ,\nn\\[3mm]
\frac{Dv^\mu}{dt}&\equiv&\ \dot v^\mu \ + \ \Gamma^\mu_{\rho\sigma}\dot x^\rho v^\sigma \ = \
\dot x^\rho D_\rho v^\mu\, .
\eea
In particular note that their commutator is
\be
\Big[\, {\cal D},\frac{D}{dt}\, \Big]v^\mu =\delta x^\rho \dot x^\sigma R_{\rho\sigma}{}^\mu{}_\nu v^\nu
%+ \Big({\cal D} \dot x^\sigma - \frac{D \delta x^\sigma}{Dt}\Big) D_\sigma v^\mu \, ,
\, .
\ee
To determine whether the variations~\eqn{susies} are symmetries
%is similar to the ${\cal N}=1$ case
is straightforward upon noting that 
$\delta (v_\mu w^\mu) = v_\mu {\cal D} w^\mu +({\cal D} v^\mu) w_\mu$.
For the simplest ${\cal N}=1$ version of the model with $(2p,Q)=(0,1)$, the
four point coupling to the Riemann tensor in the action~\eqn{action} is absent by virtue of the 
first Bianchi identity
\be
R_{[\mu\nu\rho]\sigma}=0\, ,
\ee 
and terms proportional to three $X$'s (fermions) cancel for the same reason.
Therefore the $(0,1)$ model is invariant in any background $M$.
For higher cases the terms in the action varying into three $X$'s cancel
by a conspiracy between the four and two-point couplings of these fields.
There remain variations proportional to five $X$'s. For the ${\cal N}=2$
model with $(2p,Q)=(0,2)$, these cancel by the second Bianchi identity
\be
D_{[\kappa}R_{\mu\nu]\rho\sigma}=0\, ,
\ee
so this model is also invariant in any background. So far we have just
recovered existing results. For general values of $(2p,Q)$ the obstruction
to terms quintic in $X_\alpha^\mu$ is
\be
\delta S = \frac i8 \int dt \ X^\mu\!\cdot\! X^\nu\  X^\rho\!\cdot\! X^\sigma  D_\kappa R_{\mu\nu\rho\sigma}
X^\kappa\!\cdot\! \varepsilon\, .
\ee
%where we abbreivate $-X_\alpha J^{\beta\alpha} Y_\beta \equiv X\cdot Y$. 
Hence, this theory is supersymmetric on 
manifolds of covariantly constant  curvature
\be
D_{\kappa}R_{\mu\nu\rho\sigma}=0\, ,\label{symmetric}
\ee
or in other words, locally symmetric spaces. 
 
\section{Quantization}

\label{quantum} 
 
\subsection{Canonical Analysis}

To analyze the canonical structure of the theory
it is best to work in Darboux coordinates which
are found by introducing the vielbein $e_\mu{}^m$ and writing
the action in terms of $X^m_\alpha\equiv e_\mu{}^m X_\alpha^\mu$
\be
S=\frac12\int dt \left\{\dot x^\mu g_{\mu\nu} \dot x^\nu-iX^m \cdot \frac{DX_m}{dt}+\frac14 R_{mnrs}\ 
X^m\cdot X^n\ 
X^r\cdot X^s
\right\}\, ,
\ee
where $DV_m/dt\equiv \dot V_m+\dot x^\mu \omega_{\mu m}{}^nV_n$.
Then we cast
this action into first order form
\be
S^{(1)}=\int dt 
\left\{
p_\mu \dot x^\mu - \frac i2 X_m\cdot \dot X^m
-\frac 12 \pi_\mu g^{\mu\nu} \pi_\nu
+\frac{1}{8}R_{mnrs}\ 
X^m\cdot X^n\ 
X^r\cdot X^s
\right\}\, ,
\ee
where the covariant canonical momentum is defined by
\be
\pi_\mu\equiv p_\mu+\frac i2\omega_{\mu mn} X^m \cdot X^n\, .
\ee
Since the symplectic form is canonical we immediately quantize by demanding
the operator relations\footnote{Unitarity of our quantum mechanical model 
requires signature $(\eta_{mn})={\rm diag}(+1,\ldots,+1)$, but for the study of algebras
of differential-geometric operators on semi-Riemannian manifolds this requirement can be relaxed.
Needless to say, therefore, all our results are germane to an arbitrary choice of signature.}
\be
[p_\mu,x^\nu]=-i\delta^\nu_\mu\, ,\qquad
[X^m_\alpha,X^n_\beta\}=J_{\alpha\beta}
\eta^{mn}\, .
\ee
The quantum Hamiltonian is
\bea
H&=&\ \frac12\ \pi_m \pi^m  \  -\ \  \frac18R_{mnrs} X^m\cdot X^n \, X^r\cdot X^s\nn\\[2mm]
&+&\frac i2 \omega_{mn}{}^n \pi^m -\ \  \frac1{16} \delta_{2p+Q,1} R
% -\ \ \frac18 R_{mn} X^m\cdot X^n\ -\ \frac18 R
\, ,
\eea
where we have made a definite choice of operator ordering 
reflected by the terms proportional to a naked spin connection  and the scalar curvature.
These ensure that the operator $H$ is central\footnote{To be sure, throughout this paper 
we refer to an operator as being central, when it commutes with
all the conserved charges of the underlying quantum mechanical system. The
prototypical example being the Hamiltonian.} and can be geometrically
interpreted as a generalized Laplace operator\footnote{The term proportional
to the scalar curvature vanishes for all models save $(2p,Q)=(0,1)$.
Whenever the Riemann
tensor is covariantly constant, multiplication  by the scalar curvature is
obviously a central operation. However, for the $osp(0,1)$, ${\cal N}=1$ supersymmetric model
this term 
ensures that $H$ is central in arbitrary Riemannian backgrounds.}. 
Note that we employ the operator ordering~$\pi_m \equiv e^\mu{}_m \pi_\mu$.

\subsection{States}

To study the Hilbert space, we view the operators $X^\mu_\alpha$ as bosonic and
fermionic oscillators
\be
X^m_\alpha=
\left\{
\begin{array}{cc}
\Big(a_{1,\ldots, p}^{m\dagger}, a_{1,\ldots, p}^m, 
b_{1,\ldots, q}^{m\dagger},b_{1,\ldots, q}^m\Big)\, , &
 Q =2q \mbox{ even}\, ,\\[3mm]
\Big(a_{1,\ldots, p}^{m\dagger}, a_{1,\ldots, p}^m, 
b_{1,\ldots, q}^{m\dagger},b_{1,\ldots, q}^m,b^m\Big)\, , &
 Q =2q +1\mbox{ odd,}
\end{array}
\right.
\ee 
with
$$
[a_i^m,a^\dagger_{jn}]=\delta_{ij}\delta^m_n\, ,\qquad
\{b_a^m,b_{bn}{}^\dagger\}=\delta_{ab}\delta^m_n\, ,\\[2mm]
$$
\be
\{b^m,b^n\}=\eta^{mn}\, .
\ee
The definition of the
vacuum state depends on whether $ Q $ is even or odd. In the even case, $Q = 2q$
we introduce the Fock vacuum
\be
a_i^\mu|0\rangle=0=b_a^\mu|0\rangle\, .
\ee
Excited states then correspond to
``multi-symmetric tensor--multi forms''
$$
|\Phi_{\mu_1^1\ldots\mu^1_{s_1},\cdots,\mu_1^p\ldots\mu^p_{s_p};\nu_1^1\ldots\nu^1_{k_1},\cdots,\nu_1^q\ldots\nu^q_{k_q}}(x)\rangle\ =\ \qquad\qquad\qquad\qquad\qquad
$$
$$
\Phi_{\mu_1^1\ldots\mu^1_{s_1},\cdots,\mu_1^p\ldots\mu^p_{s_p};\nu_1^1\ldots\nu^1_{k_1},\cdots,\nu_1^q\ldots\nu^q_{k_q}}\times\qquad\qquad$$
\be
\qquad\qquad\qquad\quad
a_1^{\mu_1^1\dagger}\ldots a_1^{\mu^1_{s_1}\!\dagger}
\cdots a_p^{\mu_1^p\dagger}\ldots a_p^{\mu_{s_p}^p\!\dagger}
b_1^{\nu_1^1\dagger}\ldots b_1^{\nu^1_{k_1}\!\dagger}
\cdots b_q^{\nu_1^q\dagger}\ldots b_q^{\nu_{k_q}^q\!\dagger}
|0\rangle\, .\label{evenstate}
\ee
When $p=0$ and $q=1$ these are simply differential forms, while for $p=1$, $q=0$ they
are just totally symmetric tensors. Moreover, although tensors of the above symmetry
type are not irreducible $so(d)$ representations, any tensor field on the target space
manifold can be represented this way. Although, these formul\ae\ look complicated, we will
rarely need them, and instead can rely on our supersymmetric quantum mechanical system
to provide dynamics and a simple operator algebra on spinning space.
%to symmetric tensor-valued differential forms
%\be
%|\Phi_{\mu_1\ldots\mu_s,\nu_1\ldots\nu_k}(x)\rangle=
%\Phi_{\mu_1\ldots\mu_s,\nu_1\ldots\nu_k}
%a^{\mu_1\dagger}\cdots a^{\mu_s\dagger}
%b^{\nu_1\dagger}\cdots b^{\nu_k\dagger}|0\rangle\, .
%\ee

When $ Q = 2q+1 $ is odd, we need to consider vacua for the algebra $\{b^m,b^n\}=\delta^{mn}$
which are now degenerate. As the bilinear operators $b^{[m}b^{n]}$ obey the Lorentz/rotation
algebra, vacuum states $|R\rangle$ are labeled by a representation~$R$  of~$so(d)$.
Moreover, since $\gamma^m\equiv {\sqrt{2}}\, b^m$ obey the Dirac gamma matrix algebra,
$R$ must be a spinor representation $|\alpha\rangle$, {\it i.e.},
\be
b^m |\alpha\rangle=\frac{1}{\sqrt{2}}\gamma^{m\alpha}{}_\beta|\beta\rangle\, .
\ee
 The precise choice of representation
labeled by the spinor index $\alpha$ (not to be confused with the orthosymplectic
indices carrying the same name) depends on the dimensionality of the target space.
Requiring
\be
a_i^\mu|\alpha\rangle=0=b_a^\mu|\alpha\rangle\, ,
\ee
excited states are then
``multi-symmetric tensor-spinor--multi forms''
$$
|\Phi_{\mu_1^1\ldots\mu^1_{s_1},\cdots,\mu_1^p\ldots\mu^p_{s_p};\nu_1^1\ldots\nu^1_{k_1},\cdots,\nu_1^q\ldots\nu^q_{k_q}}^\alpha(x)\rangle\ =\ \qquad\qquad\qquad\qquad\qquad
$$
$$
\Phi_{\mu_1^1\ldots\mu^1_{s_1},\cdots,\mu_1^p\ldots\mu^p_{s_p};\nu_1^1\ldots\nu^1_{k_1},\cdots,\nu_1^q\ldots\nu^q_{k_q}}\times\qquad\qquad$$
\be
\qquad\qquad\qquad\quad
a_1^{\mu_1^1\dagger}\ldots a_1^{\mu^1_{s_1}\!\dagger}
\cdots a_p^{\mu_1^p\dagger}\ldots a_p^{\mu_{s_p}^p\!\dagger}
b_1^{\nu_1^1\dagger}\ldots b_1^{\nu^1_{k_1}\!\dagger}
\cdots b_q^{\nu_1^q\dagger}\ldots b_q^{\nu_{k_q}^q\!\dagger}
|\alpha\rangle\, .\label{oddstate}
\ee
%\be
%|\Phi_{\mu_1\ldots\mu_s,\nu_1\ldots\nu_k}^\alpha(x)\rangle=
%\Phi_{\mu_1\ldots\mu_s,\nu_1\ldots\nu_k}
%a^{\mu_1\dagger}\cdots a^{\mu_s\dagger}
%b^{\nu_1\dagger}\cdots b^{\nu_k\dagger}|\alpha\rangle\, .
%\ee
Let us denote states such as~\eqn{evenstate} and~\eqn{oddstate} simply by $|\Phi\rangle$.

In both the $ Q $ even and odd cases, by demanding the vacuum to be translation invariant
\be
p_\mu|0\rangle=0=p_\mu|\alpha\rangle\, ,
\ee
we find  that the operator $i\pi_\mu$ corresponds to the covariant derivative
\be
i\pi_\mu|\Phi\rangle=
D_\mu|\Phi\rangle\, .\label{covpi}
\ee
Similarly, the first two terms of the Hamiltonian are proportional to the Laplacian
$\Delta=D_\mu D^\mu$ so that
\be
H|\Phi\rangle=\Big(-\frac12\Delta -\frac18R_{mnrs} X^m\cdot X^n \, X^r\cdot X^s
-\frac1{16}\delta_{2p+Q,1} R
\Big)|\Phi\rangle\, . 
\ee
Note that in this notation it is  important to distinguish between states $D_\mu |\Phi\rangle$ and 
$|D_\mu \Phi\rangle$. For example, when $(2p,Q)=(0,2)$,  the state $|\omega_{\mu\nu}\rangle=b_1^{\mu\dagger} b_1^{\nu\dagger}\omega_{\mu\nu}|0\rangle$ is a
two-form, while $|D_\rho\omega_{\mu\nu}\rangle=b_1^{\mu\dagger} b_1^{\nu\dagger} b_1^{\rho\dagger} \partial_\mu \omega_{\nu\rho}|0\rangle$ is its three-form exterior derivative which does not equal $D_\mu|\omega_{\nu\rho}\rangle = (D_\mu \omega_{\nu\rho}) b_1^{\nu\dagger} b_1^{\rho\dagger}|0\rangle$. In fact, it is best to think of $|\, \cdot\, \rangle$ as a machine  which takes as
input any tensor from the target space manifold and outputs  states such as~\eqn{evenstate}
or~\eqn{oddstate}.

\subsection{Conserved Charges}

To analyze the algebra of conserved charges corresponding to the symmetries\footnote{We do not consider possible additional symmetries that could arise for backgrounds with special geometries.
An excellent starting point for this important investigation are the  spinning particle
studies~\cite{Rietdijk:1989qa,Gibbons:1993ap}.}~\eqn{internal}
and~\eqn{susies},  we note
%\be
%\pi_m=e^\mu{}_m \pi_\mu
%\ee
%which obey
the identities
\bea
{} [\pi_m,\pi_n] &=&\frac12 R_{mnrs} X^r\cdot X^s +2i \omega_{[mn]}{}^r \pi_r\, ,\nn\\[1mm]
{}\ \ \ [\pi_m,X^n_\alpha]\!&=&i\omega_{m}{}^n{}_r X^r_\alpha\, ,\nn\\[1mm]
{}\ [\pi_m,x^\mu] &=&-ie^\mu{}_m\, .
\eea
Also notice that the operators
\be
M^{mn}\equiv -X^{[m}\cdot X^{n]}\, ,
\ee
generate the rotation algebra of the tangent space
\be
[M^{mn},M^{rs}]=M^{ms}\eta^{nr}-M^{ns}\eta^{mr}+M^{nr}\eta^{ms}-M^{ms}\eta^{ns}.
\ee
The conserved charges for time translations, the generalized ``supersymmetries''~\eqn{susies},
and internal symmetries~\eqn{internal} are readily computed and respectively given by
\bea
%H&=&\frac12 \pi_m \pi^m -\frac i2 \omega_n{}^{nr}\pi_r-\frac18 R_{mnrs} X^m\cdot X^n X^r\cdot X^s\, ,%\nn\\[1mm]
H\ &=& \frac12\ \pi_m \pi^m \  \  -\ \  \frac18R_{mnrs} X^m\cdot X^n \, X^r\cdot X^s\nn\\[1mm]
&+&\frac i2 \omega_{mn }{}^n \pi^m -\ \ \frac1{16}\delta_{2p+Q,1}R
\, ,\nn\\[5mm]
v_\alpha\ \!  &=& iX^m_\alpha \pi_m\, , \nn\\[3mm]
f_{\alpha\beta}&=& X^m_{(\alpha} \eta_{mn} X^n_{\beta]}\, .
\eea
Again, these are quantum results whose orderings are important (in particular the 
terms on the second line
of the Hamiltonian $H$ are higher order in $\hbar$ and are not needed for classical Poisson
brackets). 
Checking that these charges commute with the Hamiltonian, is tedious but straightforward 
using the above identities.
%at the level of Poisson brackets the term linear in~$\pi_m$
%in the Hamiltonian $H$ should be dropped. This quantum ordering term can be computed
%by requiring $H$ to  commute with the generators $v_\alpha$ but can also be more
%easily discovered by requiring that the operator $\frac12\pi_m\pi^m$ acts as the Laplacian
%$-\frac12\Delta$ on states viewed as target space tensor fields.

When the $osp(2p|Q)$ fundamental superindex $\alpha$ is odd, the operators $v_\alpha$ are ``standard'' supersymmetry
generators rotating bosons $x^\mu$ into fermions, and fermions $X^m_\alpha$ into
momenta times bosons. Otherwise, they generate novel bosonic symmetries, which 
geometrically correspond to gradient and divergence-like operations.
To compute their algebra, it is useful to note further identities
\bea
%X^r\cdot X^s&=&-X^s\cdot X^r\, ,\nn\\[1mm]
[\pi_m,\varphi^\#]\ &=&-iD_m\varphi^\#\, ,\nn\\[2mm]
[X^m_\alpha,\varphi^\#]&=&-2\varphi^m{}_s X^s_\alpha\, ,\nn\\[2mm]
[v_\alpha,\varphi^\#]\ &=&X^m_\alpha D_m\varphi^\#+2i
\varphi_{rs}X^r_\alpha \pi^s\, ,
%[v_\alpha,X^m\cdot X^n]&=&-2\pi^{[m}X^{n]}_\alpha+2i\omega^{[mn]}{}_t X^t_\alpha
%-2iX^l_\alpha \omega_l{}^{[m}{}_t X^{n]}\cdot X^t\, .
\eea
where $\varphi_{rs}(x)=-\varphi_{sr}(x)$ and is otherwise arbitrary.
Also, we denote the contraction of $X^m\cdot X^n$ on an antisymmetric 
tensors by $\#$, so 
\be
\varphi^\#\equiv\varphi_{rs} X^r\cdot X^s\, .
\ee
% and the first of 
%these
%identities relies on having equal numbers of bosonic and 
%fermionic oscillators.
It also helps to note that
\bea
[v_\alpha,\pi_n]\ &=&\frac i2 X^m_\alpha R_{mn}^\# -X^m_\alpha \omega_{mnr} \pi^r\, , \nn\\[3mm]
{}[v_\alpha,X^n_\beta\}&=&i J_{\alpha\beta} \pi^n -\omega_m{}^n{}_r X^m_\alpha X^r_\beta\, .
\eea
Detailed  yet standard computations yield the commutators between the Hamiltonian
%can be rewritten
%simply as
%\be
%H=\frac{1}{2p+Q} v\cdot v -\frac12 (p+Q/2-1) R^{\#\#}\, ,
%\ee
%and
and the generators $v_\alpha$, $f_{\alpha\beta}$ which 
{\it on any Riemannian manifold} obey the 
superalgebra:
\begin{center}
\shabox{
\begin{tabular}{c}
$[v_\alpha,v_\beta\}\ \, = J_{\alpha\beta} \Delta -\frac12 X^m_\alpha X^n_\beta R_{mn}^\#\, ,$\\[7mm]
$[f_{\alpha\beta},v_\gamma\}\, =2v_{(\alpha}J_{\beta]\gamma}\, ,$\\[7mm]
$[f_{\alpha\beta},f_{\gamma\delta}\} =4J_{(\beta(\gamma}f_{\alpha]\delta]}\, ,$\\[7mm]
$[\square,f_{\alpha\beta}]=0\, .$
\end{tabular}
}
\end{center}
\be
\label{superalgebra}
\ee
%\bea
%{}[v_\alpha,v_\beta\}\ \, &=&\!\!J_{\alpha\beta} \Delta -\frac12 X^m_\alpha X^n_\beta R_{mn}^\#
%-\frac14 J_{\alpha\beta} R^{\#\#}
%\, ,\nn\\[3mm]
%{}[f_{\alpha\beta},v_\gamma\}\, &=&2v_{(\alpha}J_{\beta]\gamma}\, ,\nn\\[3mm]
%{}[f_{\alpha\beta},f_{\gamma\delta}\} &=&4J_{(\beta(\gamma}f_{\alpha]\delta]}\, ,\nn\\[3mm]
%{}[\square,f_{\alpha\beta}]&=&0\, .
%\label{superalgebra}
%\eea
In anticipation of its {\it r\^ole} in geometry as the Laplacian, we have defined the operator
\be
\Delta \equiv -2H - \frac14 R^{\#\#} -\frac18 \delta_{2p+Q,1} R \, . 
\ee
Importantly the ``Lichnerowicz wave operator''

\begin{center}
\shabox{
$
\square\equiv -2H
$
}\end{center}
\be
\ee
is central,
{\it i.e.}, 
\begin{center}
\shabox{
$[\square,v_\alpha]=0\, ,$}
\end{center}
\be
\label{center}
\ee {\it in symmetric space backgrounds.}
Moreover, we denote 
``supersymmetrization'' with unit weight over a pair of superindices by $(\cdot\,\cdot]$.
%The computation of the first supercommutator relies 
%on the symmetric space condition~\eqn{symmetric} on the Riemann tensor but the other two
%relations hold on any manifold.
Since the  right hand side of the $[v,v\}$--super\-commutator is $J_{\alpha\beta}\square + (\mbox{curvatures})$, the above relations
constitute a finite dimensional super Lie algebra in flat backgrounds. This algebra is the central
result of the paper, because it implies a beautiful operator superalgebra on arbitrary tensors
on any manifold $M$.
 Its analysis and geometric interpretation
is the subject of the next section.

\section{Geometry} 

\label{geometry}

\subsection{Dirac Operators and ${\cal N}=1$ Supersymmetry}

In general, for $Q$ odd, the  fermionic operator 
\be
v_{2p+Q} = b^m  \pi_m =\frac{1}{\sqrt{2}}\slashed{D}\, ,
\ee
acts on states as the Dirac operator.
Moreover, at $(2p,Q)=(0,1)$, there are no bosonic oscillators and a single fermionic
one $b^m=\frac1{\sqrt{2}}\gamma^m$ corresponding to the Dirac matrices, so we have conserved charges $H$ and $v_1=\frac1{\sqrt{2}}\slashed{D}$
and no $f_{\alpha\beta}$'s as the internal symmetry group $osp(0|1)$ is empty.
Their algebra is just the ${\cal N}=1$ supersymmetry algebra
\be
\slashed{D}^2=-2H=\Delta+\frac1{4} R \equiv \square\, ,
\ee
or in terms of geometry, the Weitzenbock identity for the square of the
Dirac operator in curved space. Since the Laplace-like operator $\square$ commutes with 
the Dirac operator $\slashed{D}$, we may view it as a Lichnerowicz wave operator.
%The scalar curvature term is a second order correction in Planck's constant, so the
%classical algebra is the ${\cal N}=1$ superalgebra (we could equally well define the
%quantum Hamiltonian by the right hand side to ensure $Q^2\sim H$, but we prefer
%the current notation for its connection with geometry).

%\be
%S=\frac12\int dt \left\{\dot x^\mu g_{\mu\nu} \dot x^\nu+i\psi^\mu\, \frac{D\psi_\mu}{Dt}\right\}\, ,
%\ee
%is invariant under supersymmetry transformations
%\bea
%\delta x^\mu &=& i \psi^\mu \varepsilon \, ,
%\nn\\[2mm]
%{\cal D} \psi^\mu &=&\ \dot x^\mu \varepsilon\, .
%\eea
%for arbitrary variations while for supersymmetry transformations only the curvature term survives.
%Since $\delta (v_\mu w^\mu) = v_\mu {\cal D} w^\mu +({\cal D} v^\mu) w_\mu$ the action varies
%into $\delta S=-\frac 12 \int dt \psi_\mu \psi^\rho \varepsilon \dot x^\sigma R_{\rho\sigma}{}^\mu{}_\nu 
%\psi^\nu$ which vanishes by virtue of the Bianchi identity
%\be
%R_{\sigma[\rho\mu\nu]}=0\, .
%\ee
%The supercharge of the ${\cal N}=1$ theory 
%corresponds to the Dirac operator
%\be
%\slashed{D} = \gamma^\mu D_\mu 
%\ee
%where the Dirac matrices $\gamma^\mu$ are identified with the spinor fields $\psi^\mu$.

\subsection{Differential Forms and ${\cal N}=2$ Supersymmetry}

At $(2p,Q)=(0,2)$, we have a pair of fermionic oscillators $(b^{m\dagger},b^m)$. 
Identifying $X_1^\mu=b^{\mu \dagger}=dx^\mu$ with anticommuting coordinate differentials $dx^\mu$,
the operator 
\be
v_1 = \d\, ,
\ee
the exterior derivative, while states are differential forms. Then we identify
\be
v_2={\bm \delta}\, ,
\ee
with the codifferential and the algebra~\eqn{superalgebra} gives
\be
\{ \d, {\bm \delta} \} = -2H  = \Delta+\frac 14 R^{\#\#} \equiv \square  \, .
\ee
The right hand side, $\square = D_\mu D^\mu+\frac 14 R^{\#\#}$ equals the {\it form} Laplacian,
which is the central modification of the usual Laplace operator. 
Again, we may view it as a Lichnerowicz wave operator acting on differential forms.
%Importantly, the operator 
%$H=-\frac12 \Delta -\18 R^{\#\#}$ 
%is central. 
The internal $osp(0|2)\cong so(1,1)$
algebra has a single generator
\be
f_{12}=b^{m\dagger}b_m - \frac 12 {\rm dim}_M\, .
\ee
The operator
\be
\N\equiv b^{m\dagger}b_m
\ee
acts on states (= forms) by counting their degree (or number of indices).
Hence, we have an $u(1|1)$ superalgebra
$$
\{ \d,{\bm \delta}\} = \square\, ,\qquad [\square,\d]=0=[\square,{\bm \delta}]\, ,
$$
\be
[\N,\d]=\d\, ,\qquad [\N,{\bm \delta}]=-{\bm \delta}\, ,\qquad [\N,\Delta]=0\, .\label{u11}
\ee
Alternately, the $u(1|1)$ superalgebra of the ${\cal N}=2$ real multiplet model
can be reinterpreted as a central extension by $\square$ of the semidirect product of 
superalgebras $osp(0|2)\ltimes {\mathbb R}^{0|2}$,
where $f_{\alpha\beta}\equiv J_{\alpha\beta} \N$ generates the
 $osp(0|2)$ factor.

\subsection{Symmetric Algebras and Quantum Mechanics}

The above
discussion relating the $osp(0|1)$ and $osp(0|2)$ models to geometry
reviewed known results. We now turn to the purely bosonic\footnote{Spinning particle models
with ``commuting-spinor'' coordinates have appeared before in the literature~\cite{Deriglazov:1998ha}.}
$osp(2|0)$ model, 
and begin our presentation of new material.

Lichnerowicz introduced a modified
Laplacian $\Delta^{\!(n)}$ acting on $n$-index totally symmetric tensors~\cite{Lichnerowicz:1961}
in  order to 
facilitate a computation of the massive spin~2 propagator 
on spaces of constant curvature\footnote{Our Ricci curvature 
convention is $R_{\mu\nu}=R_{\rho\mu\nu}{}^\rho$.}
\be
R_{\mu\nu}{}^{\rho\sigma}=-\frac{2R}{d(d-1)}\, \delta^\rho_{[\mu}\delta^\sigma_{\nu]}\, .
\label{const}
\ee
(We shall employ units $R=-d(d-1)$ whenever working with constant curvature spaces. Factors
of $R$ can be reinstated by inserting appropriate powers of $1=-\frac{R}{d(d-1)}$
on the grounds of dimensionality. All formul\ae\ obtained this way are valid for spaces of positive
or negative scalar curvature. The physical cosmological constant, $\Lambda=-R/d$.) 
Low lying examples of this Lichnerowicz wave operator are
\bea
\Delta^{\!(0)}\varphi\ \  \ &=&\ \ \ \ \Delta\varphi\, ,\nn\\[2mm]
\Delta^{\!(1)}\varphi_\mu\ \ &=&\ \ (\Delta-d+1)\varphi_\mu\, ,\nn\\[2mm]
\Delta^{\!(2)}\varphi_{\mu\nu}\ &=&\ (\Delta-2d)\varphi_{\mu\nu}
+2g_{\mu\nu}\varphi_\rho{}^\rho\, ,\nn\\[2mm]
\Delta^{\!(3)}\varphi_{\mu\nu\rho} &=& (\Delta-3d-3) \varphi_{\mu\nu\rho} 
 +6 g_{(\mu\nu}\varphi_{\rho)\sigma}{}^\sigma\, , \nn\\[1mm]
&\vdots&
\eea
Formul\ae  \ for general $n$ are known (see~\cite{Hallowell} and 
also~\cite{Christensen:1978md,Warner:1982fs}) but are also a direct consequence of
the  models given here, in particular the generalization to an arbitrary symmetric space is
\be
\Delta^{\!(n)}\varphi_{\mu_1\ldots\mu_n}=\Delta\varphi_{\mu_1\ldots\mu_n}
+n(n-1)R_{(\mu_1}{}^\rho{}_{\mu_2}{}^\sigma \varphi_{\mu_3\ldots\mu_n)\rho\sigma}
+nR_{\rho(\mu_1}\varphi_{\mu_2\ldots\mu_n)}{}^\rho\, .
\ee
Here, $\Delta=D_\mu D^\mu$  is the usual Laplacian, the key point being that~$\Delta^{\!(n)}$ ``commutes'' with symmetrized trace, metric, divergence and gradient
operations
\bea
g^{\mu\nu}\Delta^{\!(n)}\varphi_{\mu\nu\rho_3\ldots\rho_n}\ \ &=&\ \ \ \Delta^{\!(n-2)}\varphi^\mu{}_{\mu\rho_3\ldots\rho_n} \, ,\nn\\[3mm]
\Delta^{\!(n+2)}g_{(\rho_1\rho_2}\varphi_{\rho_3\ldots\rho_{n+2})}&=&
g_{(\rho_1\rho_2}\Delta^{\!(n)}\varphi_{\rho_3\ldots\rho_{n+2})}\, ,\nn\\[3mm]
D^{\mu}\Delta^{\!(n)}\varphi_{\mu\rho_2\ldots\rho_n}\ \ \ &=&\ \Delta^{\!(n-1)}D^\mu\varphi_{\mu\rho_2\ldots\rho_n} \, ,\nn\\[3mm]
\Delta^{\!(n+1)}D_{(\rho_1}\varphi_{\rho_2\ldots\rho_{n+1})}&=&\ \ 
D_{(\rho_1}\Delta^{\!(n)}\varphi_{\rho_2\ldots\rho_{n+1})}\, .\label{comm}
\eea
Indeed, the Lichnerowicz wave operator $\Delta^{(n)}$ corresponds to
$-2H$, the Hamiltonian of our quantum mechanical system at $(2p,Q)=(2,0)$. 
To see this relation we recall the symmetric algebra formalism of~\cite{Hallowell} where
the operators $({\bf N, tr, g, div, grad},\square)$ acting on symmetric tensors were defined:

\vspace{.2cm}
\begin{tabular}{c|c|c}
\hline
{\it Index}&{ ${\bf N}=dx^\mu \partial_\mu$} &
Counts indices. $\phantom{S^{S^{S^S}}}\!\!$\\[2mm]
{\it Trace}&{ ${\bf tr}=g^{\mu\nu}\partial_\mu\partial_\nu$} &
Traces
a pair of indices.\\[2mm]
{\it Metric}&{ ${\bf g}=g_{\mu\nu}dx^\mu dx^\nu$}  &
Multiplies by the metric and symmetrizes.\\[2mm]
{\it Divergence }&{${\bf div}=g^{\mu\nu}\partial_\mu D_\nu$} &
The covariant divergence.\\[2mm]
{\it Gradient}&{ ${\bf grad}=dx^\mu D_\mu$} &
Symmetrized covariant derivative. \\[2mm]\hline
\end{tabular}
\vspace{2mm}

\noindent
In that work, commuting symbols $dx^\mu$ and $\partial_\mu\equiv \partial/\partial(dx^\mu)$  
with algebra\footnote{We remind the reader that the dual differential $\partial_{\mu}$ corresponds
to the quantum mechanical oscillator $a_{\mu}$, so does {\it not} act on functions of coordinates such as the metric.}
\be
[\partial_\mu,dx^\nu]=\delta^\nu_\mu\, ,\label{coordalg}
\ee
were employed for index bookkeeping and the algebra operated on sums of symmetric tensors
\be
\Phi=\sum_n \varphi_{\mu_1\ldots\mu_n}dx^{(\mu_1}\cdots dx^{\mu_n)}\, .
\ee
Moreover the operators $({\bf g,N,tr})$ formed an $sp(2)$ Lie algebra with $({\bf grad, div)}$
transforming as its fundamental doublet representation. The Lichnerowicz wave operator $\square$
was given by\footnote{Note that the equivalent relation in equation (30) of~\cite{Hallowell} is quoted 
with the wrong sign convention for the cosmological constant relative to the remainder of that work.
Also, we have shifted the definition of ${\bf c}$ by the constant $d(d-4)/4$ to conform with the uniform definition~\eqn{casimir}.}
\be
\square=\Delta+{\bf c} +\frac{d(d-4)}4  
=[{\bf div,grad}]+2{\bf c} +\frac{d(d-4)}2\, ,
\ee
where the $sp(2)$ Casimir of~\eqn{casimir} equals
\be
{\bf c}=\frac12 J^{\beta\alpha}f_{\alpha\gamma}J^{\delta\gamma}f_{\delta\beta}
={\bf g\ tr}-{\bf N}({\bf N}+d-2)
-\frac{d(d-4)}4\, .\label{c}\ee
The operator~$\square$ acts on an $n$-index symmetric tensor  by the operator~$\Delta^{\!(n)}$ given above.
Importantly,~$\square$ is central -- commuting with $({\bf N, tr, g, div,grad})$
which explains equations~\eqn{comm} above.

Identifying the Fock space oscillators of the previous section as
\be
dx^\mu \leftrightarrow a^{\mu \dagger}_1\, ,\qquad\frac{\partial}{\partial (dx^\mu)} \leftrightarrow 
a_{1\mu}
\ee
establishes an isomorphism between the symmetric space algebra and our quantum mechanical
system at $(2p,Q)=(2,0)$. States $|\Phi\rangle$ correspond to symmetric tensors $\Phi = \sum_s \phi_{\mu_1 \ldots \mu_s} dx^{\mu_1} \ldots dx^{\mu_s}$ and
the dictionary for conserved charges of the model reads
\bea
\Big(f_{\alpha\beta}\Big)&\leftrightarrow&
\left(
\begin{array}{cc}
{\bf g}&{\bf N}+\frac d2\\[2mm]
{\bf N}+\frac d2 & {\bf tr}
\end{array}
\right)\nn\\[4mm]
\Big(v_{\alpha}\Big)\ &\leftrightarrow&
\ \ \left(
\begin{array}{c}
{\bf grad}\\[2mm]
{\bf div}
\end{array}
\right)\nn\\[4mm]
-2H\ &\leftrightarrow&\  \quad\quad\square\, .
\eea
Also, the inner product and adjoint operations introduced in~\cite{Hallowell} 
correspond to the inner product of quantum mechanical states.

In fact this dictionary even provides a generalization of
the symmetric algebra formalism to any symmetric space,
rather than simply spaces of constant curvature. 
For completeness, we write out explicitly the algebra~\eqn{superalgebra}
\bea
[\TR,\G]&=&4\N+2d\,, \nn\\[3mm]
  [\TR,\GRAD]=2\DIV\,, && [\DIV,\G]=2\GRAD\, ,\nn\\[3mm]
{}[\DIV,\GRAD]&=&\square-2R_{\mu\nu\rho\sigma} dx^\mu \partial^\nu dx^\rho \partial^\sigma\, ,\nn\\[3mm]
%Commutators with the index operator \N are all of the form
{}[\N,{\cal O}]&=&{\rm wt}_{\cal O}. {\cal O}\, ,
\eea
where the weights of the index operator $N$ provide a five-grading:
\be
\begin{array}{c|cccccc}
{\cal O}&\TR&\DIV&\N&\square&\GRAD&\G\\\hline
{\rm wt}_{\cal O}&-2&-1&0&0&1&2
\end{array}
\ee
The operators $\{\G,\N,\TR\}$ generate an $sl(2,{\mathbb R})=sp(2)=osp(2|0)$
Lie algebra and $(\GRAD,\DIV)$ transform as its fundamental representation.
When the curvature vanishes we obtain the Lie algebra of the Jacobi 
group (a very useful reference is~\cite{Berndt}). Otherwise the algebra is a deformation thereof.

\subsection{Lichnerowicz Dirac Operator}

\label{LichDirac}

Our next example is the $osp(2|1)$ model. It is the first model with both 
bosonic and fermionic oscillators and is distinguished among all models
by possessing a central, Lichnerowicz--Dirac operator. This operator was first uncovered
in a study of partially massless higher spin fermi fields~\cite{Deser:2001pe}.
It displays an extremely interesting ``statistical-transmutation'' property. Namely,
that although it is an odd operator (proportional to odd numbers of Dirac matrices),
it  {\it commutes} with every other operator in the algebra (fermionic or otherwise).

The $osp(2|1)$ model has a pair of bosonic oscillators, interpreted
as commuting coordinate differentials and their duals
\be
(a^{m\dagger},a_m)\cong (dx^m, \partial_m)\, ,
\ee
with algebra~\eqn{coordalg}, just as for the $osp(2|0)$ model. 
In addition there is a single vector of fermionic
oscillators corresponding to the Dirac matrices 
\be
b^m\cong \frac1{\sqrt2}\gamma^m\, ,
\ee
as for the $osp(0|1)$, ${\cal N}=1$ supersymmetric model.
States are totally symmetric spinor-valued tensors. In addition to the Dirac operator $\slashed{D}$
of the $osp(0|1)$ model, and operators $(\N,\TR,\G,\DIV,\GRAD)$ of the $osp(2|0)$ model
and their mutual Lichnerowicz operator $\square$, there are an additional pair of
operators $({\bm \gamma},{\bm \gamma}^*)$, which either add a symmetric tensor index carried 
by a Dirac matrix, or take a gamma-trace\footnote{Strictly we should write, for example, either
${\bm \gamma}|\psi_{\mu_2\dots \mu_s}\rangle=|\gamma_{\mu_1}\psi_{\mu_2\dots \mu_s}\rangle$ 
in a quantum mechanical language, or
${\bm \gamma} \psi_{\mu_1\dots \mu_{s-1}}dx^{\mu_1}\cdots dx^{\mu_{s-1}}=
\gamma_{\mu_1}\psi_{\mu_2\dots \mu_s}dx^{\mu_1}\cdots dx^{\mu_{s}}$, 
in a geometry notation. }:
\bea
{\bm \gamma}\ : \psi_{\mu_2\dots \mu_s}&\longmapsto&
\gamma_{(\mu_1}\psi_{\mu_2\dots \mu_s)}\, ,\nn\\[2mm]
{\bm \gamma}^*: \psi_{\mu_1\dots \mu_s}&\longmapsto&
s\gamma^{\mu}\psi_{\mu\mu_2\dots \mu_s}\, .
\eea
Together, $(\G,{\bm \gamma},\N,{\bm \gamma}^*,\TR)$ generate $osp(2|1)$
\be
\Big(f_{\alpha\beta}\Big)=\left(
\begin{array}{cc|c}
\G&\N+\frac d2 &\frac1{\sqrt2}{\bm \gamma} \\[4mm]
\N+\frac d2 & \TR & \frac1{\sqrt2}{\bm \gamma}^* \\[1mm] \hline
\frac1{\sqrt2}{\bm \gamma} & \frac1{\sqrt2} {\bm \gamma}^* & 0
\end{array}
\right)\, ,\quad
[f_{\alpha\beta},f_{\gamma\delta}\} =4J_{(\beta(\gamma}f_{\alpha]\delta]}\, .
\ee
This superalgebra mostly replicates the one given above for $osp(2|0)$, 
but in addition $({\bm \gamma},{\bm \gamma}^*)$ transform as an $sp(2)$ doublet: 
\bea
[\G,{\bm \gamma}]=&0&=[{\bm \gamma}^*,\TR]\, ,\nn\\[4mm]
{}[{\bm \gamma}^*,\G]=2{\bm \gamma}\, ,&& [\TR,{\bm \gamma}]=2{\bm \gamma}^*\, ,\nn\\[4mm]
{}[N,{\bm \gamma}]={\bm \gamma}\,\  , && [N,{\bm \gamma}^*]=-{\bm \gamma}^*\, ,\nn\\[4mm]
\{{\bm \gamma},{\bm \gamma}\}=2\G\, , && \{{\bm \gamma}^*,{\bm \gamma}^*\}=2\TR\, ,\nn\\[4mm]
\{{\bm \gamma},{\bm \gamma}^*\}&=&2\N+d\, .\eea
Writing out the  second $[f,v\}$ line of the algebra~\eqn{superalgebra} yields
(for those commutation relations not given in the previous $osp(2|0)$ section)
\bea
[{\bm \gamma}^*,\GRAD]\ =&\slashed{D}&=\ [\DIV,{\bm \gamma}]\, ,\nn\\[2mm]
{}[\slashed{D},\G]\ =&0&=\ [\slashed{D},\TR]\, ,\nn\\[3mm]
{}\{\slashed{D},{\bm \gamma}\} \ = \ 2\ \GRAD\, , && 
{}\{\slashed{D},{\bm \gamma}^*\} \ = \ 2\ \DIV \, .
\eea
Finally it remains to explicate the $[v,v\}$ relations. Since these are simplest
(and most utilizable) for the case of constant curvature, we specialize to 
\be
R_{mnrs}=2\eta_{m[r}\eta_{s]n}\, .\label{constantcurvature}
\ee
Then we find
\bea
[\slashed{D},\GRAD]&=&{\bm \gamma} \Big[\N+\frac{d-1}{2}\Big] -\G{\bm \gamma}^*\, , \nn\\[3mm]
\slashed{D}^2\ \ &=& \square  -\ {\bf c} \ - \  \ \N+{\bm \gamma}{\bm \gamma}^*  -\ \frac18d(3d-5)\, , \nn\\[3mm]
{}[\DIV,\GRAD]\! &=&
\square - 2{\bf c}  - \frac12 [\N-{\bm \gamma}{\bm \gamma}^*] -\frac18d(3d-7)\, , \nn\\[3mm]
{}[\DIV,\slashed{D}]&=& \Big[\N+\frac{d-1}2\Big]{\bm \gamma}^*-{\bm \gamma}\TR\, .
\eea
Here 
\be
{\bf c}=\G\ \TR-\N(\N+d-1)+{\bm \gamma}{\bm \gamma}^* -\frac14 d(d-2)\, ,
\ee 
is the $osp(2|1)$ Casimir and the
central Lichnerowicz wave operator 
\be
\square=\Delta+{\bf c}+\frac18d(d-3)\, .
\ee
Incidentally, introducing the supergravity-inspired operators\footnote{Recall that ${\cal N}=1$
cosmological supergravity is most simply defined in terms of a modified
covariant derivative ${\cal D}_\mu = D_\mu +\frac12 {\sqrt{-\Lambda/3}} \ \gamma_\mu$~\cite{Deser:1977uq,Townsend:1977qa}. Note also that the algebra  presented in section 7 of~\cite{Hallowell} suffers some
typographical errors amounting to interchanging $(\GRAD,\DIV)\leftrightarrow ({\cal G}{\bf rad},{\cal D}{\bf iv})$. The correct algebra is presented above.}
\be
{\cal G}{\bf rad}=\GRAD+\frac i2 {\bm \gamma}\, ,\qquad
{\cal D}{\bf iv}=\DIV-\frac i2 {\bm \gamma}^*\, ,
\ee
the above algebra simplifies to
\be
[{\cal D}{\bf iv},{\cal G}{\bf rad}]=\square - 2{\bf c} -\frac38 d(d-3)\, .
\ee

Armed with the above algebra, we can now search for further central operators,
and find a generalized Dirac operator
\be
{\cal D} = {\bm \gamma} \slashed{D} {\bm \gamma}^* - {\bm \gamma}^* \slashed{D} {\bm \gamma} \, .
\ee
This is the operator introduced in~\cite{Deser:2001pe}. Although it is clearly of odd Grassmann     
parity, it commutes with all other operators $(\G$, ${\bm \gamma}$, $\GRAD$, $\N$, $\square$, $\slashed{D}$, 
$\DIV$, ${\bm \gamma}^*$, $\TR)$ in constant curvature backgrounds. In particular, it fails to commute with $\slashed{D}$
in general symmetric spaces (in which $\square$, being the Hamiltonian of our spinning particle model,
is still central). It commutes with the  $osp(2|1)$ generators in any background
and seems to have no
 generalization to higher $osp(2p|Q)$ models with $Q$ odd. So to the best of our (current) knowledge 
its existence is a peculiarity of symmetric spinor-tensors.

\subsection{Multisymmetric Tensors}

When the oscillator algebra is $osp(2p|0)=sp(2p)$, states are tensors
whose indices are grouped into totally symmetric subsets
\be
\varphi_{\mu^1_1\ldots\mu^1_{s_1},\ldots,\mu_1^p\ldots\mu_{s_p}^p}=
\varphi_{(\mu^1_1\ldots\mu^1_{s_1}),\ldots,(\mu_1^p\ldots\mu_{s_p}^p)}\, .\label{symm}
\ee
The oscillators can be viewed as commuting coordinate differentials and their duals
\be
X^\mu_\alpha=\{
dx^\mu_1,\ldots dx^\mu_p,\partial^\mu_1,\ldots,\partial^\mu_p
\}\, ,
\ee
with algebra
\be
[\partial_{i\mu},dx_j^\nu]=\delta_{ij}\delta_\mu^\nu\, .
\ee
In Young diagram notation we could depict the tensor in~\eqn{symm} as
\be
\Yvcentermath1 
\begin{array}{cc}\yng(4)& s_1 \mbox{ boxes}\\ \otimes & \\ 
  \yng(5) & s_2 \mbox{ boxes}\\ \otimes  & \\[-1mm]   \vdots &\\ \otimes \\ \yng(3)& 
  s_p \mbox{ boxes\, .}\end{array} \label{young}
\ee
The operations $\G$, $\N$ and $\TR$ of the above sections correspond to adding a pair of
boxes, counting the number of boxes, or removing a pair of boxes in the single row case, respectively.
In addition we would now like to count the number of indices in a given row, move boxes from one row to
another, and add or remove pairs of boxes from distinct rows.
All these operations are achieved by promoting $\G$, $\N$ and $\TR$ to $p\times p$
matrices of operators:
\bea
\G\equiv&\!\!(\G_{ij})\!\!&=(dx^\mu_i g_{\mu\nu}dx^\nu_j)\, ,\nn\\[2mm]
\N\equiv&\!\!(\N_{ij})\!\!&=(dx^\mu_i \partial_{j\mu})\, ,\nn\\[2mm]
\TR\equiv&\!\!(\TR_{ij})\!\!&=(\partial_{i\mu} g^{\mu\nu}\partial_{j\nu})\, .
\eea
These operators correspond precisely to the $p(2p+1)$ conserved $osp(2p|0)$ charges
of the underlying quantum mechanical model
\be
(f_{\alpha\beta})=
\left(\begin{array}{cc}
{\bf g}&{\bf N}+\frac d2{\bf 1}\\[2mm]
{\bf N}^t+\frac d2{\bf 1} & {\bf tr}
\end{array}
\right)\, ,
\ee
where ${\bf 1}$ is the $p\times p$ identity matrix and the matrices $\G$ and $\TR$
are symmetric.
These obey the $sp(2p)$ Lie algebra following from~\eqn{superalgebra}
\bea
[\N_{ij},\G_{kl}]\, &=&\; \; \; 2\delta_{j(k}\G_{l)i}\, ,\nn\\[2mm]
[\TR_{ij},\G_{kl}]\, &=&\;\, \ 4 \delta_{(k(i}\N_{l)j)}+2d\delta_{i(k}\delta_{l)j}\, ,\nn\\[2mm]
[\N_{ij},\TR_{kl}] &=&-2\delta_{i(k}\TR_{l)j}\, ,\nn\\[2mm]
[\N_{ij},\N_{kl}] &=& \; \; \; \ \delta_{jk} \N_{il} - \delta_{il} \N_{kj}
\eea 
whose
quadratic Casimir~\eqn{casimir} now reads
\bea
{\bf c}&=&{\rm tr}\Big[\frac{\G\ \TR+\TR\ \G}{2}-\Big(\N+\frac d2 {\bf 1}\Big)^2 \Big]\nn\\[4mm]&=&
\G_{ij}\TR_{ij}-\N_{ij}\N_{ji}-(d-p-1)\N_{ii}-\frac14 dp(d-2p-2)\, .
\eea
There are now gradient and divergence operators acting on each row
\be
(v_\alpha)=\left(\begin{array}{c}\GRAD_i\\[2mm]\DIV_i\end{array}\right)\, ,
\ee
which transform as the fundamental representation of $sp(2p)$
\bea
[\N_{ij},\GRAD_k]=\delta_{jk}\GRAD_i\, ,\   && [\N_{ij},\DIV_k]=-\delta_{ik}\DIV_j\, ,\nn\\[2mm]
[\TR_{ij},\GRAD_k]=2\delta_{k(i}\DIV_{j)}\, , && [\G_{ij},\DIV_k]=-2\delta_{k(i}\GRAD_{j)}\, .
\eea
They obey the algebra
\bea
[\GRAD_i,\GRAD_j]&=& R_{\mu\nu\rho\sigma} 
dx_i^\mu dx_j^\nu dx_k^\rho \partial_k^\sigma\, ,\nn\\[3mm] 
{}[\DIV_i,\GRAD_j]\ &=&\delta_{ij}\Delta - R_{\mu\nu\rho\sigma} 
dx_j^\mu \partial_i^\nu dx_k^\rho \partial_k^\sigma  \, ,\nn\\[3mm]
[\DIV_i,\DIV_j]\ \ &=& R_{\mu\nu\rho\sigma} 
\partial_i^\mu \partial_j^\nu dx_k^\rho \partial_k^\sigma \, .\label{alg1}
\eea
Moreover, the Lichnerowicz wave
operator
\be
\square = \Delta + R_{\mu\nu\rho\sigma} 
dx_i^\mu \partial_i^\nu dx_k^\rho \partial_k^\sigma \, ,
\label{alg2}
\ee
commutes with $\G$, $\N$ and $\TR$ on any manifold, and in the case of
symmetric spaces also commutes with $\DIV$ and $\GRAD$.
Finally, for constant curvature manifolds~\eqn{const}, the relations~\eqn{alg1} and~\eqn{alg2}
simplify to 
\bea
[\GRAD_i,\GRAD_j]&=&\ \ 2\G_{k[i}\N_{j]k}\, ,\nn\\[4mm]
[\DIV_i,\GRAD_j]\ &=&\delta_{ij}(\square-{\bf c} -\frac14dp[d-2p-2])\nn\\
&&\;\qquad\, - \ \G_{jk}\TR_{ki}+\N_{jk}\N_{ki}+(d-p-1)\N_{ji}\, ,\nn\\[2mm]
[\DIV_i,\DIV_j]\ \ &=&- 2\N_{[j|k|}\TR_{i]k}\, ,
\eea
and
\be
\square\  = \ \Delta +\  {\bf c} +\frac14 dp(d-2p-2)\, .
\ee

\subsection{Multiforms}

Models with $osp(0|2q)$ oscillators have been studied extensively in the literature.
The oscillator modes are all fermionic and their bilinears represent the $so(2q)$
Lie algebra. In terms of geometry, they correspond to~$q$ sets of anticommuting coordinate
differentials and their duals $\{(dx^\mu_1,\partial_{1\nu}),\ldots,(dx^\mu_q,\partial_{q\nu})\}$
subject to
\be
\{dx_a^\mu,\partial_{b\nu}\}=\delta^\mu_\nu \delta_{ab}\, ,
\ee
with indices $a,b,c,\ldots=1,\ldots, q$.
States are therefore multiforms, {\it i.e.}, tensors with groups of totally antisymmetric
indices
\be
\phi_{\nu^1_1\ldots\nu^1_{k_1},\ldots,\nu_1^q\ldots\nu_{k_q}^q}=
\phi_{[\nu^1_1\ldots\nu^1_{k_1}],\ldots,[\nu_1^q\ldots\nu_{k_q}^q]}\, .
\ee
Pictorially, the analog of~\eqn{young} would be products of columns.
Geometric operations on beasts of this type have been studied in detail~\cite{deMedeiros:2003dc,Senovilla,Bekaert:2006ix}
in the context of mixed symmetry higher spin fields. Gauged versions of the underlying 
$SO(2q)$ spinning particle models and their relation to higher spins appeared in~\cite{Kuzenko:1995mg,Bastianelli:2005vk,Bastianelli:2005uy}. 

Once again the conserved charges $f_{\alpha\beta}$, $v_\alpha$ and $H$ correspond to 
geometric operators. The $so(2q)$ generators,
\be
\Big(f_{\alpha\beta}\Big)=
\left(
\begin{array}{cc}
\G&{\bf N}-\frac{d}2 {\bf 1}\\[2mm]
-{\bf N}^t+\frac{d}2 {\bf 1} & \TR
\end{array}
\right)\, ,
\ee
either count indices from a  set or move an
index from one set to another via \be{\bf N}=(b^{m\dagger}_a \eta_{mn} b^n_b)\, ,\ee or add or remove a pair of indices from a pair of antisymmetric
index sets using the $q\times q$ antisymmetric matrices of operators
 \be\G=(b^{m\dagger}_a \eta_{mn} b^{n\dagger}_b)=-\G^t \;\;\mbox{ or }\;\;
\TR=(b^{m}_a \eta_{mn} b^{n}_b)=-\TR^t\, .\ee
Naturally, it is not possible to trace or add a pair of boxes using the metric on a single column.
As an example of these operations,
the Riemann tensor $R_{\mu\nu\rho\sigma}=R_{[\mu\nu]_1[\rho\sigma]_2}$ is a biform 
and
\be
\Big(f_{\alpha\beta}\Big)R_{\mu\nu\rho\sigma}=
\left(
\begin{array}{cccc}
0&g_{[\kappa[\eta}R_{\mu\nu]_1\rho\sigma]_2}&
    [{\ss 2-\frac d2}]R_{\mu\nu\rho\sigma}&2R_{[\mu\nu\rho]_1\sigma}\\[3mm]
-g_{[\kappa[\eta}R_{\mu\nu]_1\rho\sigma]_2}&0&
    -2R_{\mu[\nu\rho\sigma]_2}&[{\ss 2-\frac d2}]R_{\mu\nu\rho\sigma}\\[3mm]
-[{\ss 2-\frac d2}]R_{\mu\nu\rho\sigma}&2R_{\mu[\nu\rho\sigma]_2}&0&-4R_{\mu\rho}\\[3mm]
-2R_{[\mu\nu\rho]_1\sigma}&-[{\ss 2-\frac d2}]R_{\mu\nu\rho\sigma}&4R_{\mu\rho}&0
\end{array}
\right)\, ,
\ee
(for clarity we have labeled index sets by subscripts $1,2$ and have not imposed the first
Bianchi identity). Clearly, an index notation rapidly becomes cumbersome, and it is best just to 
think of the operators acting on states, but it should at least make the interpretation of these
operators clear. Their $so(2q)$ algebra follows from~\eqn{superalgebra}
\be
[f_{\alpha\beta},f_{\gamma\delta}]=4J_{[\beta[\gamma}f_{\alpha]\delta]}\, ,
\ee 
or spelled out
\bea
[\N_{ab},\G_{cd}]\, &=&\; \; \; 2\G_{a[d}{ \delta}_{c]b}\, ,\nn\\[2mm]
[\TR_{ab},\G_{cd}]\, &=&\;\, \ 4 \delta_{[c[a}\N_{d]b]}+2d\delta_{a[d}\delta_{c]b}\, ,\nn\\[2mm]
[\N_{ab},\TR_{cd}] &=&-2\TR_{b[d}\delta_{c]a}\, ,\nn\\[2mm]
[\N_{ab}, \N_{cd}] &=& \; \; \; \ \, \delta_{bc} \N_{ad} - \delta_{ad} \N_{cb}\, .
\eea 
Their Casimir~\eqn{casimir} is
\be
{\bf c}=\G_{ab}\TR_{ba}+\N_{ab}\N_{ba}-(d+q-1)\N_{aa}+\frac14dq(d+2q-2)\, .
\ee
The operators $v_\alpha$ generalize the exterior derivative and codifferential
\be
(v_\alpha)=\left(\begin{array}{c}\d_1\\[-1mm] \vdots \\ \d_q \\[1mm] \bm{\delta}_1\\[-1mm]  \vdots \\ \bm{\delta} _q\end{array}\right)\, .
\ee
These act on each antisymmetric set of indices much like the usual $\d$ and $\bm{\delta}$
operators, but it is important to note that even the operators $\d_a\cong dx_a^\mu D_\mu$
are not metric independent, since they employ the covariant derivative. The $v_\alpha$ transform as a vector under\footnote{We apologize for ambiguous notations such as:
(i) the Kronecker delta $\delta_{ab}$ versus codifferentials, $\bm{\delta}_a$, (ii) the spacetime dimension $d={\rm dim}_M$ versus
exterior derivatives $\d_a$, and (iii) the use of $\alpha$ as both an $osp(2p|Q)$
superindex and as a spacetime spinor index.}  $so(2q)$
\bea
[\N_{ab},\d_c]=\delta_{bc}\d_a\, , \!\quad && [\N_{ab},\bm{\delta}_c]=-\delta_{ac}\bm{\delta}_b\, ,\nn\\[2mm]
[\TR_{ab},\d_c]=2\delta_{c[b}\bm{\delta}_{a]}\, , && [\G_{ab},\bm{\delta}_c]=2\delta_{c[b}\d_{a]}\, .
\eea
They are supercharges from the spinning particle model viewpoint and
obey the superalgebra~\eqn{superalgebra}. 
In detail, these read
\bea
\{\d_a,\d_b\}&=&R_{\mu\nu\rho\sigma}dx_a^\mu dx^\nu_b dx_c^\rho\partial_{c}^{\sigma}  \, ,\nn\\[3mm]
\{\d_a,\bm{\delta}_b\}&=&\delta_{ab}\Delta + R_{\mu\nu\rho\sigma}dx^\mu_a \partial^\nu_b
dx^\rho_c \partial^\sigma_c\, ,\nn\\[3mm]
\{\bm{\delta}_a,\bm{\delta}_b\}&=&R_{\mu\nu\rho\sigma}\partial_a^\mu \partial^\nu_b dx_c^\rho\partial_{c}^{\sigma} 
\, .
\eea
Let us spell those relations out explicitly for the constant curvature case~\eqn{constantcurvature}
\bea
\{\d_a,\d_b\}&=&2 \G_{c(a}{\bf N}_{b)c}\, ,
\nn\\[3mm]
\{\d_a,\bm{\delta}_b\}&=&\delta_{ab}(\square-{\bf c}+\frac14 dq(d+2q-2))\nn\\&&\qquad\ \ 
                                    +\ \G_{ac}\TR_{cb}+\N_{ac}\N_{cb}-(d+q-1)\N_{ab}\, ,\nn\\[3mm]
\{\bm{\delta}_a,\bm{\delta}_b\}&=& 2\N_{c(a}\TR_{b)c}\, ,
\eea
where the Laplacian is related to the Lichnerowicz wave operator by
\be
\square=\Delta+{\bf c}-\frac14qd(d+2q-2)\, .
\ee

\subsection{Super-Lichnerowicz Algebras}

\label{superlich}

It is now clear that our supersymmetric quantum mechanical system provides a vast
generalization of Lichnerowicz's original construction. 
%The first generalization was anticipated
%in~\cite{}. Namely, that acting on symmetric spinor-tensors $\Psi^\alpha_{\mu_1\ldots\mu_s}$, 
%there is also a central ``Lichnerowicz Dirac operator'' which commutes with the operators 
%$({\bf div, grad, g, tr, N},\square)$ along with gamma and gamma-trace operations 
%$(\gamma,\gamma^*)$ that multiply by a Dirac matrix and symmetrize or take a gamma-trace, 
%respectively.
The most general extension acts on tensors with $p$ sets of symmetrized indices, 
$[Q/2]$~sets of anti-symmetrized indices and a single spinor index {\it present only for $Q$ odd}
\be
\Phi^{\alpha}_{(\mu_1\ldots \mu_{s_1}),\cdots,(\mu_1\ldots\mu_{s_p});
[\nu_1\ldots \nu_{k_1}],\cdots,[\nu_1\ldots\nu_{k_{\tiny [Q/2]}}]}\, .
\ee
In Young diagram notation, where rows are totally symmetric and columns antisymmetric,  we could write
\be
\Phi^\alpha_{\tiny\Yvcentermath1 
\!\!\!\!\!\!\!\!\!\!\!\!\!\!\!\!\!\!\mbox{$p$ times}
\left\{\!\!\!
\begin{array}{c}\yng(4)\\ \otimes \\ \yng(5)\\ \otimes \\[-1mm] \vdots \\ \otimes \\ \yng(3)\end{array}
\right.
\otimes \ \
\underbrace{\yng(1,1,1,1)\otimes\yng(1,1,1)\otimes\cdots\otimes\yng(1,1,1,1,1)}_{\mbox{$[Q/2]$ times}}}\, .
\label{tensors}
\ee
Clearly, although this is not an irreducible basis for tensors and spinors on a manifold, we can
generate all such objects this way. Indeed, the study of irreducible tensors
amounts to an examination of irreducible representations of the non-compact 
orthosymplectic algebra obeyed by the conserved charges. A detailed 
study is reserved for future work, however.

Firstly, let us discuss the $osp(2p|Q)$ generators. For any $Q$, these can (i) add a pair of boxes,
(ii) count or move boxes and (iii) remove a pair of boxes using $\G$, $\N$ and $\TR$.
These are now operator-valued supermatrices. Their fermionic entries act on pairs of boxes
of which one lives in an (antisymmetric) column, and the other in a (symmetric) row.
Otherwise, they act exactly as in the above multisymmetric tensor and multiform examples.
When $Q$ is odd, states are also labeled by a spinor index, so a single box can be added or removed
using a Dirac matrix as described in section~\ref{LichDirac}. All these operators fit into the conserved 
charges $f_{\alpha\beta}$ according to
\be
\Big(f_{\alpha\beta}\Big)=
\left(
\begin{array}{cc||cc|c}
\G_{ij}&\N_{ij}+\frac d2\delta_{ij}&\G_{ib}&\N_{ib}&\frac1{\sqrt{2}}{\bm \gamma}_i\\[5mm]
\N_{ji}+\frac d2 \delta_{ij}&\TR_{ij}&\N_{bi}&\TR_{ib}&\frac1{\sqrt{2}}{\bm \gamma}^*_i\\[2mm]
\hline\hline 
\G_{aj}&\N_{aj}&\G_{ab}&
\hspace{-1cm}\phantom{A^{A^{A^{A^A}}}}\N_{ab}-\frac d2\delta_{ab}&\frac1{\sqrt{2}}{\bm \gamma}_a\\[5mm]
\N_{ja}&\TR_{aj}&-\N_{ba}+\frac d2\delta_{ab}&\TR_{ab}&\frac1{\sqrt{2}} {\bm \gamma}^*_a \\[2mm]\hline
\frac1{\sqrt{2}} {\bm \gamma}_j&\frac1{\sqrt{2}} {\bm \gamma}_j^*&-\frac1{\sqrt{2}} {\bm \gamma}_b&
-\frac1{\sqrt{2}} {\bm \gamma}_b^*&0\phantom{0^{0^{0^{0^0}}}}\hspace{-.8cm}
\end{array}
\right)\, .
\ee
{\it The final column and row should be omitted for $Q$ even.}
Here $i,j,\ldots$ and $a,b,\ldots$ take values $1,\ldots,p$ and $1,\ldots,[Q/2]$.
The operator matrix entries are defined by
$$
\G_{\bullet\circ} \equiv dx_{\bullet}^\mu g_{\mu\nu} dx^\nu_\circ\, , \quad
\N_{\bullet\circ} \equiv dx_{\bullet}^\mu g_{\mu\nu} \partial^\nu_\circ \, ,\quad
\TR_{\bullet\circ} \equiv \partial_{\bullet}^\mu g_{\mu\nu} \partial^\nu_\circ\, ,
$$
\be
{\bm \gamma}_\bullet\equiv dx^\mu_\bullet \gamma_\mu\, ,\qquad
{\bm \gamma}^*_\bullet\equiv \partial^\mu_\bullet \gamma_\mu\, .
\ee
where $\bullet$ and $\circ$ stand for indices of either type $i,j,\ldots$ or $a,b,\ldots$.
The differentials and their duals are either commuting or anticommuting with non-vanishing
brackets
\be
[\partial_i^\mu,dx_j^\nu]=\delta_{ij} g^{\mu\nu}\, ,\qquad
\{\partial_a^\mu,dx_b^\nu\}=\delta_{ab}g^{\mu\nu}\, .
\ee
In terms of the oscillators the correspondence is $X_\alpha^\mu=(dx_i^\mu,\partial_i^\mu,
dx_a^\mu,\partial_a^\mu,\frac1{\sqrt{2}} \gamma^\mu)$.
The superalgebra of the $osp(2p|Q)$ generators $f_{\alpha\beta}$
is given in~\eqn{superalgebra} and their quadratic Casimir by~\eqn{casimir}.

Our discussion so far holds for any Riemannian manifold. 
It remains to discuss the charges $v_\alpha=i X^\mu_\alpha \pi_\mu$
and the Lichnerowicz wave operator $\square=-2H$. 
These operators can distinguish between general backgrounds, symmetric spaces, and
constant curvature ones.
The operator $i \pi_\mu$
corresponds to the covariant derivative operator on states (see equation~\eqn{covpi}).
For $\alpha=a$ odd, these are ``standard'' supersymmetry operators. From a geometric viewpoint, 
this means that they act on columns ({\it i.e.}, form indices) as exterior derivative and codifferential
operators. It is important to realize that the covariant
derivative $D_\mu$ acting on a tensor-valued differential form $\varphi_{[\mu_1\ldots\mu_k]\nu_1\ldots\nu_s}$
(say) also mixes the tensor indices $\nu_i$ through the Christoffel symbols. 
When $Q$ is odd, the final generator $v_{2p+Q}$ is the Dirac operator, as discussed in 
section~\ref{LichDirac}. Finally for values of the superindex $\alpha=i$ even, the $v_\alpha$
are gradient and divergence operators acting on symmetric rows.
In an equation, $v_\alpha$ is a column vector 
\be
(v_\alpha)=
\left(
\begin{array}{c}
{\bf grad}
\\
{\bf div}
\\
\hline
{\bf d}
\\
{\bm\delta}
\\
{\bf \slashed D}
\end{array}
\right)\, ,
\ee
which transforms as the fundamental representation of $osp(2p|Q)$ as given in~\eqn{superalgebra}.
Here, omit the final entry from $(v_\alpha)$ when $Q$ is even.
The Lichnerowicz wave operator\footnote{It is interesting to note that $\square$ can always
be expressed
in terms of the $v_\alpha$ by contracting the $[v_\alpha,v_\beta\}$ bracket with $J^{\alpha\beta}$.
Moreover, when $Q\neq 2p$ an even simpler formula 
$$
\square=\frac{1}{p-Q/2}\ v\cdot v +\frac14 \frac{p-Q/2+1}{p-Q/2} R^{\#\#} +\frac18\delta_{2p+Q,1}R\, .
$$
holds upon contracting with $J^{\beta\alpha}$.
}
is again a modification of the Laplacian according to
\be
\square=\Delta+\frac 14 R^{\#\#}+\frac18 \delta_{2p+Q,1} R\, .
\ee
It commutes with $f_{\alpha\beta}$ in any background and with $v_\alpha$ in
symmetric spaces.
The algebra of all the above geometric operations is given in~\eqn{superalgebra}
and is valid for any Riemannian manifold.
We propose to call this a``super-Lichnerowicz algebra''.
The additional relation~\eqn{center}, required for $\square$ to be central always holds
for $(2p|Q)=(0|1,2)$ (${\cal N}=1,2$ supersymmetry) and is otherwise valid in symmetric 
spaces~\eqn{symmetric}.

We end this section with its specialization to constant curvature manifolds~\eqn{const}.
In that case the operator
\be
R^{\#\#}=4{\bf c}-\frac12 d(Q-2p)(Q-2p+d-2) \, , 
\ee
{\it i.e.},  the $osp(2p|Q)$ quadratic Casimir. The supercommutators of the $v_\alpha$
then simplify to
\be
[v_\alpha,v_\beta\}=J_{\alpha\beta} \Delta + f_{[\alpha|\gamma}J^{\delta\gamma}f_{\delta|\beta)}
-\frac14d(Q-2p+d-2)J_{\alpha\beta}\, .
\ee
The Lichnerowicz wave operator is simply
\be
\square=\Delta + {\bf c} -\frac18 d(Q-2p)(Q-2p+d-2) +\frac18 \delta_{2p+Q,1}R.
\ee

\subsection{Parabolic Orthosymplectic Algebra}

\label{totalalg}

The superalgebra~\eqn{superalgebra}  is certainly consistent
since it was defined by the charges of a quantum mechanical system acting on
an explicit representation by quantum states or equivalently tensors on a manifold.
However, the presence of terms quartic in the oscillators $X_\alpha^m$ on the 
right hand side of the $[v,v\}$-supercommutator imply that this is not a closed algebra.
(The reader may also easily convince themselves of this fact by studying the constant 
curvature, $(2p,Q)=(2,0)$ case studied in depth in~\cite{Hallowell}.)
However, when the curvature vanishes, we do obtain a finite dimensional Lie superalgebra.
As we now show, it is a parabolic subalgebra of $osp(2p+2|Q)$.
The curved models are deformations of this parabolic algebra.

The first step is to introduce a new operator ${\bf ord}$
which counts derivatives:
\be
[{\bf ord},H]=2H\, ,\quad [{\bf ord},v_\alpha]=v_\alpha\, ,\quad [{\bf ord},f_{\alpha\beta}]=0\, .
\ee
Although ${\bf ord}$ is not a conserved charge, it can be interpreted as a dilation generator, 
and will play an extremely important {\it r\^ole}. In particular, it provides 
a 3-grading of the Lie superalgebra
\be
{\mathfrak p}={\mathfrak g}_2\oplus {\mathfrak g}_1\oplus {\mathfrak g}_0\, ,
\ee
where ${\mathfrak g}_2=\{H\}$, ${\mathfrak g}_1=\{v_{\alpha}\}$
and $\mathfrak g_0=\{f_{\alpha\beta},{\bf ord}\}$.
Moreover, $\{H,{\mathfrak g}_1,{\bf ord}\}$ form a Heisenberg Lie superalgebra
\be
[v_\alpha,v_\beta\}=-2J_{\alpha\beta} H\, .
\ee
It is natural to wonder therefore, whether the algebra ${\mathfrak p}$ is the parabolic subalgebra
of some larger Lie superalgebra, and indeed this is the case. Namely,
the Lie superalgebra $osp(2p+2|Q)$ has a five grading by the Cartan generator, which we are here calling ${\bf ord}$, corresponding
to the longest root of its bosonic $sp(2p+2)$ subalgebra
\be
osp(2p+2|Q)={\mathfrak g}_2\oplus {\mathfrak g}_1\oplus {\mathfrak g}_0\oplus {\mathfrak g}_{-1}\oplus {\mathfrak g}_{-2}\, .
\ee
The non-negatively graded subspaces are isomorphic to the parabolic algebra ${\mathfrak p}$,
while we have as yet found no physical interpretation of the negatively graded subspaces,
but cannot help remarking that the corresponding quantum mechanical generators ought be
computable in terms of inverse powers of~$H$.

In summary, the algebra of conserved charges corresponds to all $osp(2p+2|Q)$ generators
that commute with the generator $H$ labeled by the longest root of the bosonic $sp(2p+2)$
subalgebra. To verify this claim explicitly, it suffices to display an upper-triangular matrix representation
of the super Lie algebra ${\mathfrak p}$. This is achieved by the $osp(2p+2|Q)$ supermatrix
\be
{\cal P}\equiv
\left(
\begin{array}{c|c|c}
{\bf ord}&v_\alpha&4H\\[2mm]\hline
0&2J^{\beta\gamma}f_{\gamma\alpha}&-J^{\beta\gamma}v_\gamma
\phantom{A^{A^{A^{A^A}}}}\hspace{-1.1cm}
\\[2mm]\hline
0&0&-{\bf ord}\phantom{A^{A^{A^{A^A}}}}\hspace{-1cm}
\end{array}
\right)\, ,
\ee
(it is not difficult to permute the rows and columns of ${\cal P}$ to obtain the usual
$osp$-valued block bosonic/fermionic supermatix form).

\section{Conclusions}

\label{conclusions}

In this article, we have presented curved space spinning particle models 
whose spin degrees of freedom are described by both fermionic and
bosonic oscillators. These transform as spacetime vectors and
under the superalgebra $osp(2p|Q)$. States of these models describe the most general spinors
and tensors on any Riemannian manifold. The conserved charges describe the various
geometric operators on tensors and spinors. For manifolds that are symmetric spaces,
our models are supersymmetric with $Q$ supercharges corresponding to exterior derivatives,
codifferentials, and (when $Q$ is odd) the Dirac operator. Moreover, the underlying superalgebra is a deformation
of a parabolic subalgebra of $osp(2p+2|Q)$ with $2p$ ``commuting'' supercharges which
act as gradient and divergence operators. The Hamiltonian of the model is a modified Laplacian
that generalizes Lichnerowicz's constant curvature wave operator to tensors and spinors
on arbitrary Riemannian manifolds. The remaining charges amount to all possible operations
on tensors and spinors using the metric and Dirac matrices. 

There are many applications, generalizations, and open directions suggested by our work.
We close by giving a (partial) list:
\begin{enumerate}
\item In special geometries it is possible to write down 
quantum mechanical systems with higher supersymmetries~\cite{Coles:1990hr,Hull:1999ng}, 
the first example being K\"ahler 
backgrounds~\cite{Zumino:1979et}. 
It is possible motivate the super-Lichnerowicz algebra presented here
by analogy with the Hodge-Lefshetz algebra of Dolbeault cohomology which is
also described in terms of supersymmetric quantum mechanics~\cite{Figueroa-O'Farrill:1997ks}. 
It would be interesting
to consider K\"ahler backgrounds for our models  to try and marry the two algebras
(an investigation of this direction may be found in~\cite{Marcus:1994mm}).

\item We have said nothing here about the spectrum of the model or its detailed dynamics.
Eigenmodes of Lichnerowicz operators have been studied before~\cite{Christensen:1978md,Warner:1982ps}, but it would be interesting
to have statements applicable to our more general models. Clearly this is
an important problem deserving further attention.

\item An important relation of supersymmetric quantum mechanics to geometry is through index theorems. The Lichnerowicz operators presented here are natural regulators for index computations
in an extremely broad class of spaces, in particular, from a geometric standpoint the quantum 
mechanical trace over states runs over infinitely many fields. Whether novel invariants can be constructed this way is an open problem.

\item Second quantization of these models introduces fields of arbitrary spin,
thanks to the introduction of bosonic oscillators~\cite{Labastida:1987kw,Vasiliev:1988xc}. 
Already the Lichnerowicz algebra for  the $osp(2|0)$ model
implies generating functions for totally massless, symmetric, higher spin field actions. Indeed, 
calling
\bea
G&=&\square+2(\N-1)(\N+d-3)-\GRAD\ \DIV+\frac12\Big(\GRAD^2\ \TR+\G\ \DIV^2\Big)
\nn\\[1mm]
    &-&\frac12\G\Big(\square+2\N(\N+d-1)+d-1+\frac12\GRAD\ \DIV\Big)\TR\, ,
\eea
the (Bianchi) identity
\be
\DIV G = \G X = 0 \mbox{ mod } \G\, ,
\ee
implies that the action
\be
S=\int \Phi^* G \Phi = \langle \Phi|G|\Phi\rangle\, , \label{generate}
\ee
is gauge invariant under $\delta \Phi = {\GRAD}\ \xi$ when $\TR\ \xi=0$. This is the 
invariance of a  doubly trace free, $\TR^2 \Phi=0$, field. Since $\Phi$ need not be an 
eigenstate of the index operator $\N$, the action~\eqn{generate} generates the actions
of all totally symmetric, massless fields in constant curvature backgrounds.

Similarly, the $osp(2|1)$ model pertains to totally symmetric tensor-spinors. Indeed
the operator
\bea
G&=&{\cal D}-2\N\ \slashed{D}+(d-4)\Big[{\bm \gamma}\slashed{D}{\bm \gamma}^*-{\cal G}{\bf rad}\ {\bm \gamma}^*-{\bm \gamma}\ {\cal D}{\bf iv}\Big]
\nn\\[1mm]
               &+&\frac12\ (d-2)\Big[{\bm \gamma}\ {\cal G}{\bf rad}\ \TR+\G\ {\cal D}{\bf iv}\ {\bm \gamma}^*
                       -i(2\N+d-4)\Big]
\nn\\[1mm]
  &-&  \frac14\  (d-2)\Big[\G\Big(\slashed{D}-\frac i2(2\N+d)\Big)\TR-2i{\bm \gamma}(2\N+d-2){\bm \gamma}^*\Big]
\, ,\nn\\
\eea
%\begin{verbatim}
%local G1=Dirac  -2*N*Dslash-i_*(d-2)*(N+(d-4)/2) 
%
%              -1/4*(d-2)*(g*(Dslash-i_*(N+d/2))*tr-2*i_*[g]*(2*N+d-2)*[g*])
%                       
%                          +(d-4)*([g]*Dslash*[g*]-Grad*[g*]-[g]*Div)
%
%                        +1/2*(d-2)*([g]*Grad*tr+g*Div*[g*]);
%\end{verbatim}
obeys the identity
\be
{\cal D}{\bf iv} G = {\bm \gamma} X = 0 \mbox{ mod } {\bm \gamma}\, ,
\ee
so that the action
\be
S=\int \overline{\Psi} G \Psi = \langle \Psi| G |\Psi\rangle,
\ee
enjoys the gauge invariance $\delta \Psi =\overline {{\cal G}{\bf rad}}\ \xi$ with ${\bm \gamma}\xi=0$,
of a traceless, gamma-traceless, ${\bm \gamma}^* \TR\ \Psi = 0$,
fermionic higher spin field.

An obvious approach for obtaining the above actions is to gauge the rigid $osp$
symmetries~\cite{Brink:1976sz} 
of the models. Spinning particle studies of this type have been conducted 
in~\cite{Howe:1988ft,Howe:1989vn,Bastianelli,Hull:1993ay,Halling:1990iu}.
Interactions would amount to second quantization of these spinning particle models.
Needless to say, this is an ambitious program, see~\cite{Bekaert:2005vh}
for a review of interacting higher spin theories.

\item The super-Lichnerowicz algebra is a deformation of a parabolic $osp(2p+2|Q)$
algebra. It would be most interesting to find geometric operators corresponding to the remaining
$osp(2p+2|Q)$ operators. In particular, it seems likely that introducing the Green's function $
\square^{-1}$ may yield a solution to this problem. The additional charges would not be expected
to yield quantum mechanical  symmetries, but instead ought be spectrum generating.

\item The Lichnerowicz algebra for the $osp(2|0)$ model is a deformation of 
the Jacobi group $G^J$, an object extensively studied in the mathematical literature (see
the excellent book~\cite{Berndt}). In particular
it is known that this group enjoys a cubic Casimir operator, in our notations
\be
{\bf c}_3= \square \Big({\bf c}+\N+\frac{d(d-4)}{4}
\Big)
-\GRAD^2\ \TR-\G \ \DIV^2+ \GRAD\ (2\N+d-2) \ \DIV\, .
\ee
This operator is central in flat backgrounds. We know of no generalization to 
curved backgrounds however.
\item Instead of studying the model via the Lie algebra obeyed
by its conserved charges, for the $osp(2|0)$ theory it is also helpful to 
work in an extended universal enveloping algebra.
As shown in~\cite{Hallowell}, in this case there is
a rather elegant associative algebra. 
The key step is to enlarge the constant curvature
algebra by a certain square root of the Casimir
\be
\Tt\equiv-\sqrt{1-\c}
\ee
and in addition we define $\Nt\equiv \N+\frac d2-1$.
This allows us to form the operator $\Nt+\Tt$ whose eigenstates
are $k$-fold trace-free tensors, namely
\be
\TR^k \varphi =0\neq \TR^{k-1}\varphi
\ \Longrightarrow \
(\Nt+\Tt) \varphi = 2k \varphi\, .
\ee
Then introducing
\be
\wt \DIV\equiv (\Nt-\Tt)\DIV-\GRAD\ \TR, 
\ee
and similarly for the formal adjoint $\wt \GRAD$, the constant 
curvature algebra is presented by the six relations
\bea
\TR\ \Nt=(\Nt+2)\TR\,, &&
\TR\ \wt \GRAD=\wt \GRAD\ \frac{\Nt-\Tt+4}{\Nt-\Tt+2}\ \TR\, ,\nn\\[3mm]
\G\ \TR&=&\Nt^2-\Tt^2=\TR\ \G-4\Nt-4\, ,\nn\\[4mm]
\wt\DIV \Tt=(\Tt-1)\wt\DIV\, ,&&
\wt\DIV\  \Nt=(\Nt+1)\wt\DIV\, ,\nn\\[-2mm] \nn
\eea
$$
\wt\DIV\ \wt\GRAD\ =\ \wt\GRAD\ \wt\DIV\frac{(\Nt-\Tt+2)\Tt^2}{(\Nt-\Tt)(\Tt^2-1)}
$$
\be
\hspace{3.7cm}
-\ 2\frac{(\square-\frac{(d-2)^2}{2}+2\Tt^2)(\Nt-\Tt+2)\Tt^2}{\Nt-\Tt}\, ,
\label{casalg}
\ee
and their formal adjoints where all other products are commutative.
In particular, observe that the $sp(2,\Real)$ action on the pair
$(\wt\DIV,\wt\GRAD)$ is diagonal.
These relations provide a calculus for constant curvature algebra
computations in terms of rational functions of $(\Nt,\Tt)$.
Needless to say, it would be extremely fruitful to generalize this
associative algebra to the general $osp(2p|Q)$ models.

\end{enumerate}

\section*{Acknowledgments}
It is a pleasure to thank Boris Pioline, Sergey Prokushkin, and 
Andrew Hodge for discussions.
%This work was supported in part by the 
%National Science Foundation under grant PHY01-40365.

\end{document}